%% file: ms.tex
\documentclass[a4paper,11pt]{article}
\pdfoutput=1
\usepackage{jcappub}
\usepackage{bm}
\usepackage{xspace}
\usepackage[section]{placeins}
\graphicspath{{plots/}}
\renewenvironment{figure*}{\figure}{\endfigure}

\newcommand{\hMpc}{\,h\mathrm{Mpc}^{-1}}
\newcommand{\Msun}{M_\odot}
\newcommand{\dc}{\delta_{\rm sc}}
\newcommand{\dl}{\delta_{\rm b}}
\newcommand{\Mh}{M_h}\newcommand{\Mf}{M_f}\newcommand{\Ms}{M_s}
\newcommand{\bso}{b_{s^2}}
\newcommand{\bGa}{b_{\Gamma_3}}
\newcommand{\whm}{\mbox{WHM}\xspace}
\newcommand{\pbs}{\mbox{PBS}\xspace}
\newcommand{\whmpbs}{\mbox{\textsc{whm-pbs}}\xspace}
\newcommand{\nh}{n_h}\newcommand{\nf}{n_f}
\newcommand{\pd}[2]{\frac{\partial #1}{\partial #2}}

\title{Web--Halo Model Peak--Background Split (WHM-PBS): halo bias as a distribution, not a number}

\author{Samuel~Brieden and}
\author{Alexander~Tipp}

\affiliation{Institute for Theoretical Particle Physics and Cosmology (TTK), RWTH Aachen
University, Sommerfeldstr.~16, D-52056 Aachen, Germany}
\emailAdd{brieden@physik.rwth-aachen.de}
\emailAdd{alexander.tipp@rwth-aachen.de}

\abstract{
We present the Web--Halo Model Peak--Background Split (\whmpbs), an analytic theory in which the large-scale bias of a dark-matter halo is inherited from its cosmic-web environment. Building on the Web--Halo Model, we use the Shen \textit{et al.} moving barriers for ellipsoidal collapse generating the web hierarchy in which every halo sits inside a host filament, itself inside a sheet. Combined with the peak--background split, this picture replaces the deterministic bias--mass relation $b(M_h)$ with the bias of the host environment, averaged over the conditional mass function. As a result, halo bias $b(M_h)$ is no longer a number but a strongly skewed \emph{distribution}. In this work we make use of this distribution in three different ways: as (i) a physically motivated prior on bias relations, (ii) a prediction on halo stochasticity, and (iii) a framework for assembly bias models. Regarding (i) we find that the density bias relations $b_2(b_1)$ and $b_3(b_1)$ stay tight, while the tidal bias $b_{s^2}(b_1)$ shows significant scatter, as found in $N$-body simulations.
Regarding (ii), once including halo exclusion, our model reproduces the super- to sub-Poisson shot-noise trend of Baldauf \textit{et al.} which we convert into a prior band on the EFT stochasticity amplitude parameters. Finally, regarding (iii) in the density sector it explains the bias--concentration--correlation inversion of Paranjape \textit{et al.} at the characteristic mass ($M_\mathrm{h}\simeq1.7\times10^{13}\,h^{-1}\Msun$), with no parameter tuned to assembly bias.
In the tidal sector, we use the measurements from Lazeyras \textit{et al.} to calibrate the tidal bias prior of (i), tying the three uses together.
We apply the resulting priors on synthetic data, demonstrating that the \whmpbs priors mitigate projection effects arising when all nuisance parameters are varied freely, while fixed priors may fail severely. The code to reproduce our results is publicly available at \href{https://github.com/SamuelBrieden/whmpbs.git}{https://github.com/SamuelBrieden/whmpbs.git}.
}

\begin{document}
\maketitle
\flushbottom

\input{sections/intro}
\input{sections/excursion_set}
\input{sections/moving_barrier_bias}

\input{sections/predictions}

\input{sections/stochasticity}

\input{sections/assembly_bias}

\input{sections/demo}
\input{sections/conclusions}

\acknowledgments
We would like to thank Fabian Schmidt, Andrew Zentner, Julien Lesgourgues, Markus Mosbech, Antonia Mattes and Markus Braun for useful conversations. We acknowledge that parts of this work were carried out with the help of agentic AI (Anthropic's Claude Code). Specifically, we used it to implement from scratch the Zhang-Hui solver, refine the numerical bias calculation, and draft sections 5-8, and appendices A-D. We used it as an assistant to generate from scratch the plotting scripts for figures 1, 7, 8, 10, 11, 12, 13, and 14, while we also used it to refine all remaining figures, whose scripts pre-existed before the adoption of Claude Code, but making use of the truncated series. Also Sections 1-4 were drafted before its adoption. We verified each step of the pipeline and take full responsibility for the scientific content of this work. The numerical implementation of the model and the scripts reproducing all figures are publicly available at \href{https://github.com/SamuelBrieden/whmpbs.git}{https://github.com/SamuelBrieden/whmpbs.git}.

\appendix
\input{sections/appendices}

\bibliographystyle{JHEP}
\bibliography{refs}

\end{document}

%% file: sections/intro.tex
\newpage
\section{Introduction}
\label{sec:intro}
% [2026-08-04 field-level round, shortened 2026-08-05] Two novelty claims re-scoped, because the
% analytic Lagrangian bias-field literature (arXiv:2604.00097 and refs. therein) predates ours on
% arXiv (31 Mar vs 23 Jul 2026):
%   F2 (para 1): the clause claiming the bias distribution had 'thus far only' been SEEN in N-body
%       simulations is replaced by 'so far studied mainly in N-body simulations and in
%       complementary Lagrangian frameworks' + a bulk citation.
%   F3 (para 5): 'jointly' inserted into the first-time claim -- the JOINT bias + stochasticity +
%       scatter prediction survives that literature; a bias-distribution-alone claim would not.
% Author rule for this round: the field-level works are cited in BULK only, none singled out or
% named in prose, no constructions/equations reproduced; the related-work paragraph below keeps
% only the DIFFERENCES from this work.

The clustering of dark-matter haloes is biased relative to the matter \citep{Kaiser1984,colekaiser1989,bond1991}: on large scales the halo overdensity tracks the matter overdensity through the linear bias $b_1$, a single number that grows with halo mass. However, this is only the leading term. The full relation between a tracer and the underlying matter is a perturbative expansion in the density and tidal fields, comprising a series of deterministic bias coefficients (the linear $b_1$, the higher-order $b_2$ and $b_3$, the tidal $b_{s^2}$, and so on) together with stochastic, shot-noise-like contributions \citep{mcdonaldroy2009,desjacques2018}. But again, for each halo mass there is a \emph{fixed} set of bias numbers. Simulations tell a more interesting story. At fixed mass, haloes cluster differently according to secondary properties such as formation time, concentration, or environment, an effect known as \emph{assembly bias} \citep{gao2005,ShethTormen2004,dalal2008,wechslertinker2018}. One particularly interesting incarnation of assembly bias is the bias--concentration correlation (BCC), whose sign inverts near the characteristic mass $m_\ast\!\approx\!2\times10^{13}\,h^{-1}\Msun$ \citep{wechsler2006,gao2007,paranjape2018}: more concentrated haloes are more strongly clustered than average at low mass and less strongly clustered at high mass. Mass alone, then, does not fix a halo's bias. Instead, $b(M)$ is the \emph{mean} of a distribution, whose width is as physical as the mean itself. In this work we derive an analytical prediction for this distribution, and argue that its width, and not only its mean, is where much of the predictive power lies. 

The classic analytic route to halo bias is the peak--background split \citep[\pbs;][]{shethtormen1999,cooray2002,Schmidt2013PBS,desjacques2018}: the response of the halo abundance to a long-wavelength background mode. In the excursion-set picture \citep{press1974,bond1991} this abundance is determined from the first crossings of a barrier by an ensemble of random walks in the smoothed density field. The textbook barrier is a constant, derived from spherical collapse. But real collapse is \emph{ellipsoidal}: the threshold a patch must reach depends on the tidal shear that rises with the filtering variance, and is hence given by a \emph{moving} barrier \citep{smt2001,shethtormen2002,shen2006}. That same shear is what organises matter into the cosmic web of sheets, filaments and haloes \citep{bondweb1996}, each morphology with its own moving barrier \citep{shen2006}. The property we exploit is that the \pbs applies to \emph{any} barrier, so one machinery delivers the large-scale bias of sheets, filaments and haloes alike.

This is the ingredient on which we build the \emph{Web--Halo Model Peak--Background Split} (\whmpbs), a successor to the Web--Halo Model \citep[WHM;][]{whm}. Its single idea is that a halo does not collapse in isolation: it forms inside a filament, which is itself embedded in a sheet, and its large-scale bias is \emph{inherited} from that hierarchy of environments. Concretely, the bias of a halo is the bias of its host filament, averaged over the conditional mass function $P(\Mf|\Mh)$ that links a halo of mass $\Mh$ to its host filament of mass $\Mf$ (Section~\ref{sec:excursion}). Because the host mass can vary at fixed halo mass, the inherited bias is a distribution: its mean is the familiar large-scale bias, and its width measures the variety of environments a halo of that mass can occupy. The width depends on whether hosts are identified at the filament or the sheet level and we always show and discuss both cases.

The width of the bias distribution is the central result of this paper, and we interpret it in three different ways, leading to three different predictions for observables:
\begin{itemize}
\item[(i)] Interpreted as a \emph{selection uncertainty}, it becomes a physically motivated prior band around the bias relations that encompasses the corresponding $N$-body measurement (Section~\ref{sec:predictions}). 
\item[(ii)] Interpreted as a \emph{random field}, it sources the stochasticity of the halo power spectrum, providing a physical prior on the EFT stochasticity coefficients in agreement with simulation results (Section~\ref{sec:pk}).
\item[(iii)] Interpreted as a \emph{correlation} between environment and halo structure, it becomes assembly bias. Regarding $b_1$, from the filament host population we derive the sign inversion of the bias--concentration correlation near the characteristic mass $m_\ast$ observed in N-body simulations (Section~\ref{sec:assembly_mixing})
For $\bso$, this same correlation calibrates the selection prior of (i) (Section~\ref{sec:tidal_assembly}), so assembly bias is not merely a corollary but part of the prior set.
\end{itemize}

Predicting the bias (i) and stochasticity (ii) coefficients is of fundamental importance. They are nuisance parameters of the effective-field-theory (EFT) full-shape analyses through which surveys such as DESI \citep{desicollab2016,desidr2bao,desi2024fs,desi2024cosmo,desi2024joint} and \emph{Euclid} \citep{laureijs2011} measure the expansion history and the growth of structure. In such analyses the coefficients can be treated in one of three ways: left free under broad priors, which opens the door to prior-volume (projection) effects that can bias the inferred cosmology \citep{Hadzhiyska2023priors,carrilho2023priors,Tsedrik2026priors}; fixed to assumed values, which risks a systematic error should those values be inaccurate; or constrained by physically motivated priors. This third route has thus far only been possible by calibrating to $N$-body simulations, and priors of this kind are known to tighten full-shape constraints \citep{eggemeier2021} and be applicable to galaxies as well \citep{barreira2021,zennaro2022priors} under the assumption of a Halo Occupation Distribution (HOD) model. The aim of this work, to our knowledge for the first time, is to jointly predict halo bias and stochasticity relations \emph{and their scatter} analytically from first principles. We test our set of priors on an idealistic, synthetic dataset (Section~\ref{sec:demo}), where broad nuisance priors alone can bias the recovered cosmology through projection effects that the \whmpbs priors partially remove.

Bias as a field or a distribution, rather than a fixed set of coefficients, is the starting point of several recent approaches \citep{stuecker2025gauss,stuecker2025cumulant,maion2025shape,banerjee2026}, which reach it from the statistics of the initial conditions. Two things set the \whmpbs apart. First, the distribution here is generated by the collapse selection itself: the web barriers decide which \emph{host} a halo of mass $\Mh$ ends up in, so it is the spread of hosts, not of points, that we propagate into $P(b|\Mh)$. Second, that spread is skewed rather than Gaussian.

The paper is organised as follows. Section~\ref{sec:excursion} sets up the cosmic-web excursion set and its mass functions. Section~\ref{sec:bias} derives the environment-averaged density and tidal biases. Section~\ref{sec:predictions} compares the predicted distribution and the $b_N(b_1)$ relations against simulations and assembles the selection prior band. Section~\ref{sec:pk} develops the stochasticity
 and Section~\ref{sec:assembly} the bias--concentration inversion and the tidal-sector assembly-bias signal that calibrates the tidal prior's selection term. Section~\ref{sec:demo} demonstrates the priors in a synthetic DESI-like inference, and Section~\ref{sec:conclusions} concludes. Appendix~\ref{app:taylor} compares the exact first crossing with the truncated series that earlier versions of this work relied on; the remaining appendices describe the numerical ZH solver implementation (\ref{app:zh}), the exact density ($\delta$)-response bias (\ref{app:pbsbias}), and the exact tidal ($s^2$)-response (\ref{app:tidal}).
Throughout, we adopt a flat $\Lambda$CDM cosmology with $(\Omega_m,h,\sigma_8,n_s,\Omega_b)=(0.3096,0.6766,0.8102,0.9665,0.0490)$ and spherical collapse threshold density $\dc=1.6759$ unless specified otherwise.

%% file: sections/excursion_set.tex
\section{The cosmic-web excursion set}
\label{sec:excursion}

\subsection{The walk and the moving barriers}
\label{sec:walk}
We work in the excursion-set theory framework \citep{press1974,bond1991}. The main idea is that smoothing the linear density field on a succession of smaller scales can be modelled as a random walk: as the smoothing radius shrinks, the smoothed overdensity $\delta$ of a given patch walks up and down with the filtering variance $S=\sigma^2(M)$, which plays the role of time. We compute $\sigma^2(M)$ from the linear power spectrum with a real-space top-hat filter and the corresponding mass $M=4\pi\bar\rho_m R^3/3$, and treat the walk as Markovian: the increments of $\delta$ in $S$ are uncorrelated, so that $\delta(S)$ is a Brownian motion with $\delta(0)=0$ and $\langle\delta(S)\delta(S')\rangle=\min(S,S')$, here $S$ and $S'$ are two smoothing scales of the \emph{same} patch's walk, and $\langle\cdot\rangle$ averages over patches. This is the standard excursion-set convention \citep{bond1991,smt2001}.\footnote{The Markovian property is exact only for a filter sharp in $k$-space; the real-space top-hat introduces weak correlations between steps. Following standard practice, we neglect these.}

A patch is counted as part of a structure of a given morphology once its walk first up-crosses the relevant collapse threshold, or \emph{barrier}, $B(S)$. For spherical collapse this is just the constant critical density $B=\dc$. Real collapse, though, is \emph{ellipsoidal}: a region collapses along its three principal axes successively, and the density it must reach to do so depends on the surrounding tidal shear \citep{bondmyers1996,smt2001}. That shear lives in the $3\times3$ deformation tensor, whose six independent components would turn the one-dimensional walk into a six-dimensional problem \citep{chiueh2001,Stuecker2018rw6D}. The simpler alternative formulated by Sheth, Mo, and Tormen \citep[SMT01;][]{smt2001} (see appendix A therein) and adopted here, is to
fix the tensor's eigenvalues to their most likely values found from their probability distribution at each variance, first derived by Doroshkevich~\cite{doroshkevich1970}. This reduces the six numbers to a single effective threshold that rises with $S$ (smaller-mass, i.e.\ larger-$S$, patches feel stronger shear and so must reach a higher density to collapse): the \emph{moving} barrier
\begin{equation}
\label{eq:barrier}
B_{\rm morph}(S)=\sqrt a\,\dc\Big[1+\beta_{\rm morph}\big(S/a\dc^{2}\big)^{\alpha_{\rm morph}}\Big],
\end{equation}
whose precise form is derived from the ellipsoidal-collapse model of SMT01 and extended to the other web morphologies by Shen, Abel, Mo and Sheth~\citep[S06;][]{shen2006}.

Here, the normalisation $a=0.707$ from SMT01 is calibrated to simulations and largely reflects the halo definition, shifting with the halo finder and overdensity criterion used to identify objects in the simulations \citep{despali2016}.\footnote{It can also be interpreted as an effective parameter accounting for the fact that in excursion set theory only the density region above the threshold collapses, although it is the entire peak that should enter, which leads to a systematic underprediction of the collapsed mass.} It represents the only ``free'' parameter in this framework; everything else is fixed by the linear power spectrum and by ellipsoidal collapse. I.e., solving the ellipsoidal collapse differential equations S06 determined the fitting parameters  $(\beta,\alpha)=(0.45,0.61)$ for haloes, $(-0.012,0.28)$ for filaments and $(-0.56,0.55)$ for sheets (see appendix A therein).

Figure~\ref{fig:barriers} draws these three barriers together with a sample walk. All three share the origin $\sqrt a\,\dc$ at $S\to0$ and then fan out by morphology, with $B_{\rm sheet}<B_{\rm fil}<B_{\rm halo}$. The three barriers therefore nest, such that the walk's first up-crossings are automatically ordered: the patch first flattens into a sheet, then contracts into a filament, then collapses into a halo. The three crossings of one and the same walk \emph{are} the nested hierarchy of this paper, and reading off their scales gives the masses $\Ms>\Mf>\Mh$ of a halo and its hosts. Different walks cross at different scales, and this walk-to-walk spread is precisely what the conditional mass functions of the next subsection quantify.

\begin{figure}
\centering
\includegraphics[width=\columnwidth]{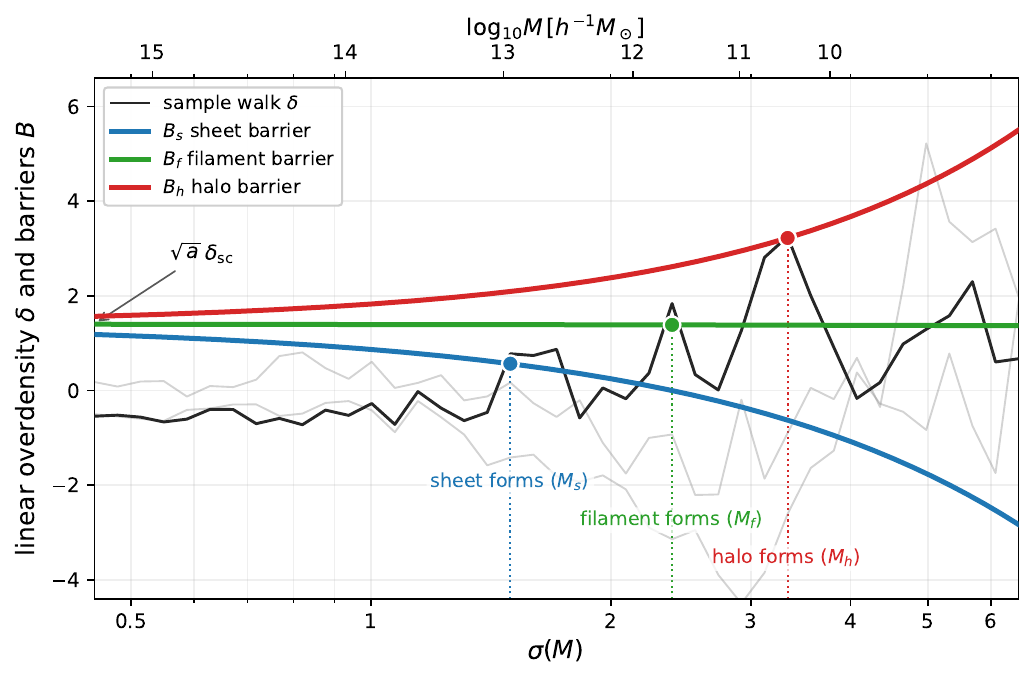}
\caption{The excursion-set skeleton of \whmpbs, in the conventions of Fig.~1 of Shen et al.~\cite{shen2006}: the moving barriers \eqref{eq:barrier} for sheets (blue), filaments (green) and haloes (red), with their parameters and $a=0.707$, against $\sigma(M)$ (logarithmic; top axis: the corresponding mass). The barriers converge to $\sqrt a\,\dc$ as $\sigma\to0$ and nest, $B_s<B_f<B_h$, at every finite $\sigma$. A sample sharp-$k$ random walk (black), sampled at scales uniform in $\log\sigma$ so that each increment carries its Brownian variance $\Delta S$, therefore first up-crosses the barriers in the fixed order sheet$\to$filament$\to$halo (dots, with dotted drop lines at the crossing scales): 
Once a sheet of mass $\Ms$ is formed at $\sigma(\Ms)$, within the same patch a filament of mass $\Mf$, and a halo of mass $\Mh$ are formed successively, at larger $\sigma$. This delivers the nested hierarchy of hosts on which this paper is built.
The two fainter walks illustrate that the crossing scales, and with them the host masses, differ from walk to walk. That spread is what the conditional mass functions of Fig.~\ref{fig:cond} quantify.}
\label{fig:barriers}
\end{figure}

\subsection{The unconditional and conditional mass functions}
\label{sec:massfunctions}
The first-crossing distribution $f(S)$ of a barrier $B(S)$ by the Markov walk is the solution of the Zhang and Hui~\cite{zhanghui2006} (ZH) Volterra integral equation,
\begin{equation}
\label{eq:fcross}
f(S)=g_1(S)+\int_0^S f(S')\,g_2(S,S')\,\mathrm dS',
\end{equation}
with kernels fixed by the barrier and its slope $B'(S)=\mathrm dB/\mathrm dS$,
\begin{equation}
\label{eq:kernels}
\begin{aligned}
g_1(S)&=\Big[\tfrac{B(S)}{S}-2B'(S)\Big]\,P_0\big(B(S),\,S\big),\\
g_2(S,S')&=\Big[2B'(S)-\tfrac{B(S)-B(S')}{S-S'}\Big]\\
&\hspace{5.5em}\times P_0\big(B(S)-B(S'),\,S-S'\big),
\end{aligned}
\end{equation}
where $P_0(\delta,s)=(2\pi s)^{-1/2}\exp(-\delta^2/2s)$ is the free Gaussian propagator. Earlier moving-barrier studies instead approximate $f(S)$ by a Taylor expansion in the barrier shape, truncated at fifth order. Its form was introduced by Sheth and Tormen~\cite{shethtormen2002} (equation (5) therein) and adopted by Shen et al.~\cite{shen2006}. Unlike this \emph{truncated series}, which earlier versions of this work also used and which we examine in Appendix~\ref{app:taylor}, the ZH equation \eqref{eq:fcross} is exact, non-negative, and normalised by construction: $\int_0^\infty f\,\mathrm dS=P_{\rm cross}$.\footnote{The total probability $P_{\rm cross}$ for any walk to cross the barrier is unity for the decreasing sheet and filament barriers, below unity for the increasing halo barrier. But in practice, one cannot reach the $S\rightarrow 0$ limit, since the ellipsoidal collapse theory eventually breaks down.} We solve it as detailed in Appendix~\ref{app:zh}. 

The unconditional mass function (abundance per volume per mass bin) of each morphology follows as
\begin{equation}
\label{eq:nM}
n(M)=\frac{\bar\rho_m}{M}\,\Big|\frac{\mathrm dS}{\mathrm dM}\Big|\,f(S),
\end{equation}
where $f(S)$ is obtained from the barriers \eqref{eq:barrier} of each morphology. For the moment we focus on the simplest non-trivial case, a halo inside its parent filament. The full halo within a filament within a sheet hierarchy is treated in Section~\ref{sec:nested}.

A halo of mass $\Mh$ forms inside a host filament of mass $\Mf>\Mh$ when its walk, having crossed the filament barrier $B_f$ at $S_f$, subsequently crosses the higher halo barrier $B_h$ at $S_h>S_f$~\citep{bower1991,laceycole1993}. The mean number of mass-$\Mh$ haloes per filament of mass $\Mf$ is given by the \emph{conditional} first crossing,
\begin{equation}
\label{eq:condmf}
N(\Mh|\Mf)=\frac{\Mf}{\Mh}\,\Big|\frac{\mathrm dS_h}{\mathrm d\Mh}\Big|\,F(S_h|S_f),
\end{equation}
where $F(S_h|S_f)$ is the solution of \eqref{eq:fcross} for the \emph{shifted} barrier $\tilde B(s)=B_h(S_f+s)-B_f(S_f)$ with $s=S_h-S_f$, corresponding to the excess threshold the walk must reach beyond its host filament crossing.
By construction, the unconditional halo and filament abundances and the conditional haloes-in-filament abundance satisfy the Chapman--Kolmogorov consistency relation \citep{bond1991}
\begin{equation}
\label{eq:CK}
\nh(\Mh)=\int N(\Mh|\Mf)\,\nf(\Mf)\,\mathrm d\Mf.
\end{equation}
The inverted conditional probability that a \emph{given} halo of mass $\Mh$ is hosted by a filament of mass $\Mf$ follows by Bayes' theorem,
\begin{equation}
\label{eq:conditional}
P(\Mf|\Mh)=\frac{N(\Mh|\Mf)\,\nf(\Mf)}{\nh(\Mh)}.
\end{equation}

The numerator integrated over $\Mf$ evaluates to exactly $\nh(\Mh)$ due to Eq.~\eqref{eq:CK}, so the probability is automatically normalised. We call this self-consistency \emph{closure}: summing the (conditional) number of haloes per filament over all host filaments returns the total (unconditional) halo abundance. As we shall see in Section~\ref{sec:envbias}, the same property ensures that host filament bias averaged over $P(\Mf|\Mh)$ returns the halo bias computed directly from $\nh(\Mh)$. The exact ZH solver adopted here automatically satisfies closure, but the truncated series does not; we compare both approaches in more detail in Appendix~\ref{app:taylor}. 

The conditional abundance $N(\Mh|\Mf)$ of Eq.~\eqref{eq:condmf} (mass-weighted, $N(\Mh|\Mf)\,\Mh/\Mf$) and its Bayesian inverse $P(\Mf|\Mh)$ [Eq.~\eqref{eq:conditional}] are shown in the left and middle panels of the top two rows of Fig.~\ref{fig:cond}, respectively. The upper row shows the 2D maps, the lower row one-dimensional slices at fixed $\Mf$ and $\Mh$. A filament of fixed mass $\Mf$ is most probable to host haloes of slightly smaller mass (strong peak of $N(\Mh|\Mf)\,\Mh/\Mf$ at $\Mh \lesssim \Mf$ in panel (d)) with a long tail towards $\Mh \ll \Mf$. Conversely, in panel (e) the probability $P(\Mf|\Mh)$ for a halo at fixed mass to reside within a filament peaks at $\Mf \gtrsim \Mh$, and the mean is dragged away from the median towards the high $\Mf$ tail. This result is consistent with figures 4 and 5 of S06.

\subsection{The nested halo-in-filament-in-sheet hierarchy}
\label{sec:nested}
The physically complete picture is the \emph{nested} sheet$\to$filament$\to$halo chain. The random walk first crossing from the sheet barrier up to the halo barrier can be broken at the first crossing of the intermediate filament barrier: the number of haloes of mass $\Mh$ forming inside a sheet of mass $\Ms$ can be found by counting the haloes inside filaments of every mass $\Mf$ hosted by the sheet, and summing over $\Mf$,
\begin{equation}
\label{eq:nest}
N(\Mh|\Ms)=\int N(\Mh|\Mf)\,N(\Mf|\Ms)\,\mathrm d\Mf .
\end{equation}
This is the Chapman--Kolmogorov composition equivalent to \eqref{eq:CK} but carried one level deeper. Equation~\eqref{eq:nest} states that the three-level hierarchy \emph{collapses} onto a single two-level conditional distribution with the sheet as the host.

The halo-in-sheet host distribution follows by Bayes' theorem exactly as the halo-in-filament one did in \eqref{eq:conditional},
\begin{equation}
\label{eq:conditional_s}
P(\Ms|\Mh)=\frac{N(\Mh|\Ms)\,n_s(\Ms)}{\nh(\Mh)},
\end{equation}
and again integrates to unity over $\Ms$. 

\begin{figure}[h]
\centering
\includegraphics[width=\textwidth,height=0.63\textheight,keepaspectratio]{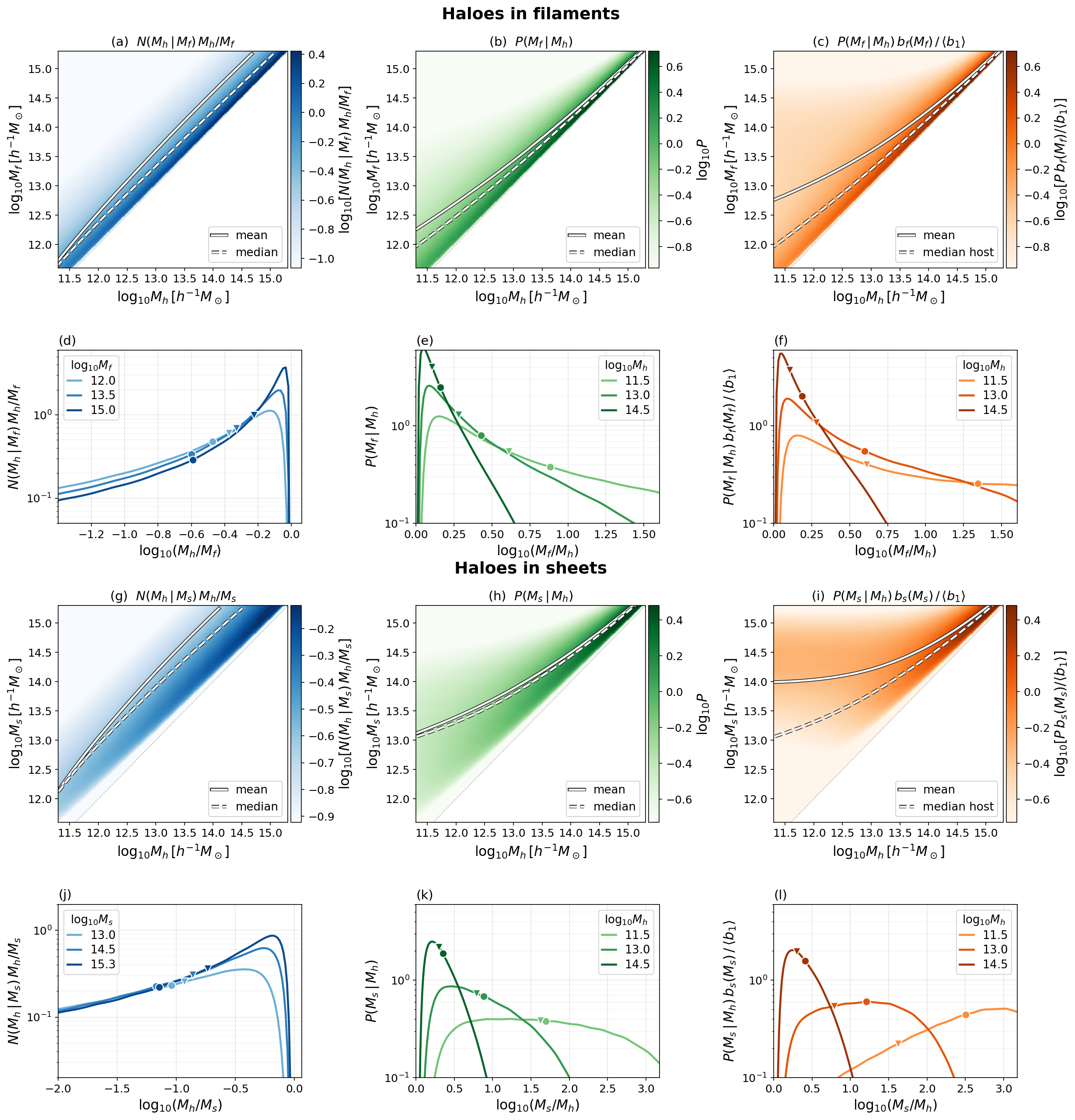}
\caption{The cosmic-web conditional mass functions and the build-up of the environment-averaged bias, from the exact ZH first crossing. \emph{Top two rows:} the two-level halo-in-filament building block; \emph{bottom two rows:} the full three-level halo-in-filament-in-sheet (nested) hierarchy, which collapses into the halo-in-sheet hierarchy due to closure. In each block the columns are [(a)/(g)] $N(M_h|M_x)$, the conditional abundance of haloes of mass $M_h$ in a host of mass $M_x$ ($x=f$ filament, $x=s$ sheet)$N(M_h|M_x)\,M_h/M_x$, the conditional abundance of Eq.~\eqref{eq:condmf} weighted by the mass ratio $M_h/M_x$ ($x=f$ filament, $x=s$ sheet) --- the fraction of the host's mass in haloes of mass $M_h$. The weighting makes it peak near the host mass, whereas the bare abundance $N(M_h|M_x)$ rises towards low $M_h$; [(b)/(h)] $P(M_x|M_h)$, its Bayesian inverse [Eqs.~\eqref{eq:conditional},~\eqref{eq:conditional_s}], the host of a halo of mass $M_h$; [(c)/(i)] and the bias-weighted host distribution $P(M_x|M_h)\,b_x(M_x)/\langle b_1\rangle_{M_x|M_h}$. The first row of each block shows the two-dimensional maps, with the mean (solid) and median (dashed) lines; the second row [(d)--(f), (j)--(l)] shows representative one-dimensional slices, each a per-dex probability density (integral over the log-mass $=1$)%
, with the mean ($\bullet$) and median ($\blacktriangledown$) marked on every curve. In panels (c)/(i) the solid track is the bias-weighted mean \emph{host mass}, which \emph{differs} between the two hosts (a halo's filament and sheet hosts are different structures of different mass), while the dashed track (labelled \emph{median host}) is the host median of (b)/(h). The host distribution at fixed halo mass is broad and right-skewed; the filament is the narrower two-level building block and the sheet the broader nested level. What closure makes identical between the two hosts is not these host-mass tracks but the \emph{integral} of (c)/(i), the mean bias $\langle b_1\rangle_{M_x|M_h}$ [Eq.~\eqref{eq:envbias}], plotted as the level-independent mean in Fig.~\ref{fig:b1}.}
\label{fig:cond}
\end{figure}

The two bottom rows of Fig.~\ref{fig:cond} show the haloes-in-sheet quantities. The first column is the conditional abundance $N(\Mh|\Ms)$, quantifying how many haloes of mass $\Mh$ a sheet of mass $\Ms$ contains mass-weighted conditional $N(\Mh|\Ms)\,\Mh/\Ms$, the fraction of a sheet's mass in haloes of mass $\Mh$. Again, as in the haloes-in-filament case, it reaches a peak and falls steeply once $\Mh$ approaches the host sheet mass. The second column inverts this via Bayes' theorem into the quantity we actually need in Section~\ref{sec:bias}: $P(\Ms|\Mh)$, the probability distribution of sheet hosts a fixed-mass halo can occupy. Again, it is broad and right-skewed, with a long tail toward the rare, massive host sheets.

%% file: sections/moving_barrier_bias.tex
\section{The environment-averaged peak--background-split bias}
\label{sec:bias}
On large scales the halo field is a local expansion in the operators of the matter density and tidal fields \citep{desjacques2018},
\begin{equation}
\label{eq:biasexp}
\delta_h = b_1\,\delta + \tfrac12 b_2\,\delta^2 + \tfrac16 b_3\,\delta^3
         + \bso\,s^2 + \bGa\,\Gamma_3 + \dots,
\end{equation}
with the density operators $\delta,\delta^2,\delta^3$ carrying the density biases $b_1,b_2,b_3$, the traceless tidal shear $s^2\equiv s_{ij}s^{ij}$ carrying the Eulerian tidal bias $\bso$ (sometimes referred to as Galileon bias $b_{\mathcal{G}_2}$), and the cubic non-local operator $\Gamma_3$
(Section~\ref{sec:bgamma3}) carrying $\bGa$. The moving-barrier first-crossing response predicts all of these as the response of the halo abundance to a long-wavelength perturbation, with no parameters beyond the barrier itself. The density biases (Section~\ref{sec:envbias}) and the tidal bias (Section~\ref{sec:tidal}) are direct first-crossing responses, computed natively in Lagrangian space and carried to Eulerian by the standard co-evolution equations. The cubic non-local $\bGa$ (Section~\ref{sec:bgamma3}), which has no Lagrangian counterpart --- $\Gamma_3$ vanishes in the initial conditions --- is generated by that same evolution as a function of $b_1$ and the Lagrangian tidal bias $\bso^L$. Note that the expansion in Eq.~\eqref{eq:biasexp} is not complete: it is missing the higher-derivative bias terms, which we leave for future work.

\subsection{Density biases \texorpdfstring{$b_N$}{bN}}
\label{sec:envbias}
In the \pbs framework a tracer $x$'s $N$-th-order Lagrangian bias is defined as the normalised response of its abundance $n_x(M)$ to a long-wavelength background perturbation $\dl$ \citep{desjacques2018,lazeyras2016},
\begin{equation}
\label{eq:pbsbias}
b_{x,N}^{L}(M)=\frac{1}{n_x(M)}\,
\frac{\partial^{N} n_x(M\,|\,\dl)}{\partial\dl^{N}}\bigg|_{\dl=0}.
\end{equation}
Such a background overdensity raises the density on all sub-scales uniformly, which is equivalent to lowering the barrier, $B(S)\to B(S)-\dl$. The conditional abundance $n_x(M\,|\,\dl)$ can hence be evaluated as the exact response of the ZH first crossing to the shifted barrier (Appendix~\ref{app:pbsbias}). Because the abundance is $n_x\propto f(S)$ up to a mass Jacobian independent of $\dl$, this normalised response is expressed equivalently in the first-crossing form
\begin{equation} 
\label{eq:pbsbias-f}
b_{x,N}^{L}(S)=\frac{1}{f(S)}\,\frac{\partial^{N}\hat f(S;\dl)}{\partial\dl^{N}}\bigg|_{\dl=0},
\qquad
\hat f(S;\dl)\equiv f\big[\,B-\dl\,\big](S),
\end{equation}
where $f[\,\cdot\,]$ denotes that the first-crossing density is a functional of the whole barrier (Appendix~\ref{app:pbsbias}) and $S=\sigma^2(M)$. This first-crossing form is what we evaluate in practice. Next, we insert the Chapman-Kolmogorov decomposition \eqref{eq:CK} into \eqref{eq:pbsbias}. A background $\dl$ lowers the halo and host barriers by the \emph{same} amount, so the barrier difference $\Delta B=B_h-B_f$, and with it the conditional crossing $N(\Mh|\Mf)$, is unchanged. The conditional crossing therefore passes straight through the $\dl$-derivative, leaving
\begin{equation}
\label{eq:envbias}
\begin{aligned}
b_{h,N}^{L}(\Mh)
&=\frac{1}{\nh(\Mh)}\int N(\Mh|\Mf)\,
\frac{\partial^{N}\nf(\Mf|\dl)}{\partial\dl^{N}}\bigg|_{0}\,\mathrm d\Mf\\
&=\int P(\Mf|\Mh)\,b_{f,N}^{L}(\Mf)\,\mathrm d\Mf \\
&\equiv \langle b_{f,N}^{L}(\Mf) \rangle_{\Mf|\Mh},
\end{aligned}
\end{equation}
where the second equality recognises the filament bias \eqref{eq:pbsbias} and the conditional host distribution \eqref{eq:conditional}. Equation~\eqref{eq:envbias} is the relation on which the remainder of the paper rests, the \emph{environment-averaged bias} (indicated via the $\langle ... \rangle_{\Mf|\Mh}$ notation): the bias of a halo is the bias of its host environment, averaged over all the hosts it can occupy.\footnote{Theoretical models of galaxy bias take a very similar form, writing the effective and higher-order galaxy bias as an integral over the host-halo probability distribution inferred from halo-occupation (HOD) models \citep{seljak2000,peacocksmith2000,benson2000,scoccimarro2001,berlindweinberg2002,cooray2002}, with Scoccimarro et al.~\cite{scoccimarro2001} giving all bias orders in exactly this host-averaged form. Recent work recasts the higher-order and tidal galaxy biases the same way and ties them to the EFT basis \citep{voivodicbarreira2021,akitsu2024,ivanov2024}.} We have written it here for the filament host, but due to \emph{closure} it holds identically for the higher level sheet host ($\Mf\!\to\!\Ms$, Section~\ref{sec:nested}). 
By definition, evaluating halo bias directly from the halo mass function \eqref{eq:pbsbias} yields the same result as the environmental average \eqref{eq:envbias}, so one may ask what we actually learn from the latter equation.
The novel aspect of this work is the \emph{scatter}: although the mean reproduces the direct bias, the host mass is random at fixed $\Mh$, so the halo bias inherits the full width of $P(\Mf|\Mh)$: \emph{it is a distribution, not a number}. The right column of Fig.~\ref{fig:cond} displays the normalised integrand of Eq.~\eqref{eq:envbias} for filament and sheet hosts, in 2D or 1D at fixed halo mass. We observe that the full host distribution differs strongly between filaments and sheets, the latter exhibiting a stronger skew towards high host masses. The impact of these differences on the predicted scatter is discussed in more detail in Section~\ref{sec:pk}. As already mentioned, we adopt the filament-only case as our default choice for the haloes' hosts. It gives a narrower bias scatter favoured by the data of Section~\ref{sec:pk}. Including sheet hosts introduces a broader scatter retained only as an upper bound. We gather the full case for this choice in Section~\ref{sec:conclusions}.  

We map from Lagrangian to Eulerian space via the co-evolution relations \citep{fry1996,mojingwhite1997,chan2012,saito2014},%
\begin{equation}
\label{eq:coevol}
\begin{aligned}
b_{h,1}^{E}&=1+ b_{h,1}^{L},\qquad
 b_{h,2}^{E}= b_{h,2}^{L}+\tfrac{8}{21} b_{h,1}^{L},\\
 b_{h,3}^{E}&= b_{h,3}^{L}-\tfrac{13}{7} b_{h,2}^{L}
-\tfrac{796}{1323} b_{h,1}^{L} .
\end{aligned}
\end{equation}
To keep the notation concise, we drop the $E$ superscripts and the $h$ subscripts when referring to Eulerian halo bias in what follows.

\subsection{Tidal bias \texorpdfstring{$\bso$}{bs2}}
\label{sec:tidal}

The environment-averaged bias \eqref{eq:envbias} was derived for the response to a
long-wavelength \emph{density}, which lowers every barrier uniformly. A long-wavelength tidal shear
$s^2\equiv s_{ij}s^{ij}$ \citep{mcdonaldroy2009,clmp1998} acts differently. Its Lagrangian bias
$\bso^{L}$ is again a first-crossing response, but to a \emph{non-uniform} barrier shift. Mirroring the density definition \eqref{eq:pbsbias}, it is the normalised response of the absolute abundance $n_x(M)$, now to a long-wavelength tidal-shear background $s_b^2$,
\begin{equation}
\label{eq:tid-def}
\bso^{L}(M)=\frac{1}{n_x(M)}\,
\pd{n_x(M\,|\,s_b^2)}{s_b^2}\bigg|_{s_b^2=0}.
\end{equation}
The shear enters through the moving part of the barrier \eqref{eq:barrier}, the $\beta$ term, which is itself the imprint of the shear, since smaller-mass (larger-$S$) patches feel more shear and need a higher threshold for collapse. Quantitatively, the barrier depends on $S$ only through the shear: for a Gaussian field the trace-free shear variance is $\langle s^2\rangle=\tfrac23 S$, so the ellipsoidal-collapse barrier is a function of the shear alone, $B=B(\langle s^2\rangle)$ \citep{smt2001}. Here $\langle s^2\rangle$ is the variance of the \emph{small-scale} shear built up along the walk. The background shear $s_b^2$ --- not resolved by the walk and absent from $S$, exactly as the density background $\dl$ in Section~\ref{sec:envbias} --- adds coherently to the shear felt at the formation scale, $\langle s^2\rangle\to\langle s^2\rangle+s_b^2$. At fixed $S$, i.e. with the small-scale density walk untouched, it raises the barrier \emph{non-uniformly} by
\begin{equation}
\label{eq:tid-shift}
\delta B(S)=\pd{B}{\langle s^2\rangle}\,s_b^2=\frac32\,\frac{\mathrm{d}B}{\mathrm{d}S}(S)\,s_b^2.
\end{equation}
In exact analogy to the density response \eqref{eq:pbsbias-f}, the unconditional tidal bias is the
normalised first-crossing response to this shift,
\begin{equation}
\label{eq:tid-uncond}
\bso^{L}(S)=\frac{1}{f(S)}\,
\frac{\partial\hat f(S;s_b^2)}{\partial s_b^2}\bigg|_{s_b^2=0},
\qquad
\hat f(S;s_b^2)\equiv f\big[\,B+\tfrac32\,s_b^2\,\mathrm{d}B/\mathrm{d}S\,\big](S),
\end{equation}
equivalently the response to shifting the barrier's shear argument, $\langle s^2\rangle\to\langle s^2\rangle+s_b^2$. Because the shift is non-uniform, this response is a genuine functional convolution of the first-crossing kernel with the barrier slope, and \emph{not} a single-scale product, see Appendix~\ref{app:tidal}. Also note that a \emph{constant} (spherical) barrier has $\mathrm{d}B/\mathrm{d}S=0$, so that $\bso^{L}=0$. Hence, the tidal bias is entirely a moving-barrier effect. 

This has an important implication for computing the full \whmpbs tidal bias distribution. 
Unlike the density background $\dl$, a long-wavelength shear $s_b^2$ does \emph{not} leave the conditional crossings invariant: it shifts each barrier by the same shear through its own slope, entering Eq.~\eqref{eq:tid-uncond}. And because the halo, filament and sheet barriers have \emph{different} slopes, the barrier differences $B_h-B_f$ and $B_f-B_s$ change, such that the conditional mass functions $N(\Mh|\Mf)$ and $N(\Mf|\Ms)$ change.%
This means that the conditionals do not pass out of the derivative as in Eq.~\eqref{eq:envbias} for the density sector, but instead each conditional response to the background shear must be kept.

We briefly sketch the consequence of this for the haloes-in-filament case, where for simplicity  we treat the filament barrier as a constant (i.e., $\mathrm{d}B_f/\mathrm{d}S=0$, which is nearly correct). Because of this, when inserting the Chapman--Kolmogorov decomposition \eqref{eq:CK} into the definition \eqref{eq:tid-def}, the host abundance $n_f(\Mf)$ passes straight through the derivative and the entire response is carried by the conditional mass function,
\begin{equation}
\label{eq:tid-envbias}
\begin{aligned}
\bso^{L}(\Mh)
&=\frac{1}{\nh(\Mh)}\int
\pd{N(\Mh|\Mf\,;s_b^2)}{s_b^2}\bigg|_{s_b^2=0}\,\nf(\Mf)\,\mathrm d\Mf\\
&=\int P(\Mf|\Mh)\,\bso^{L}(\Mh|\Mf)\,\mathrm d\Mf\\
&\equiv \langle \bso^{L}(\Mh|\Mf)\rangle_{\Mf|\Mh},
\end{aligned}
\end{equation}
where the second equality recognises the \emph{conditional} tidal bias -- the definition \eqref{eq:tid-def} applied to the conditional crossing, $\bso^{L}(\Mh|\Mf)\equiv N(\Mh|\Mf)^{-1}\,\partial N/\partial s_b^2\big|_{s_b^2=0}$ -- and the same conditional host distribution \eqref{eq:conditional} as in the density average \eqref{eq:envbias}. The tidal bias of a halo is therefore an average over the \emph{same} host distribution as its density biases, but with the weight supplied by the conditional crossing response instead of the host's. Again, the full width of $P(\Mf|\Mh)$ propagates into a distribution of $\bso$ at fixed halo mass. 

We show the explicit functional form of the exact calculation -- the small but non-zero filament and sheet slopes retained through all three web levels -- in Appendix~\ref{app:tidal}. Again, the theory \emph{closes} by construction, i.e.,
\begin{equation}
\label{eq:tid-closure}
\langle \bso^{L}(\Mh|\Mf)\rangle_{\Mf|\Mh} = \bso^{L}(\Mh)\;
\qquad\text{(as for }b_1,b_2,b_3),
\end{equation}

The Eulerian tidal bias is obtained from the local Lagrangian bias as \citep{chan2012,baldauf2012,saito2014},
\begin{equation}
\label{eq:bs2E}
\bso^{E}=\bso^{L}-\tfrac27 b_1^{L},
\end{equation}
which is non-zero also when $\bso^{L}=0$.

\subsection{Cubic non-local bias \texorpdfstring{$\bGa$}{bGamma3}}
\label{sec:bgamma3}
The same tidal response fixes another bias parameter: the cubic non-local bias $\bGa$ of \eqref{eq:biasexp}, the coefficient of the third-order Galileon-type operator $\Gamma_3=\mathcal{G}_2[\Phi]-\mathcal{G}_2[\Phi_v]$ \citep{chan2012,assassi2014,desjacques2018}. It is the difference between the second Galileon operator $\mathcal{G}_2=(\partial_i\partial_j\Phi)^2-(\nabla^2\Phi)^2$ \citep{mcdonaldroy2009,chan2012} evaluated on the gravitational potential $\Phi$ and on the velocity potential $\Phi_v$. Because $\delta=\theta$ in the initial conditions the two coincide there, such that $\Gamma_3$ and also $\bGa^L$ vanish.
The Eulerian-space-counterpart $\bGa^E \equiv \bGa$ is given by the co-evolution relation \cite{abidi2018} as
\begin{equation}
\label{eq:bgamma3}
\bGa = \tfrac{23}{42}\,(b_1-1)\;-\;\tfrac52\,\bso^{L},
\end{equation}
with the first term being the standard local-Lagrangian co-evolution value \citep{eggemeier2020} and the second one the imprint of the initial tidal field. Since our moving barrier model predicts $\bso^{L}$ (Section~\ref{sec:tidal}), Eq.~\eqref{eq:bgamma3} turns that into a (derived) prediction for $\bGa$.
We compare it against simulations in Section~\ref{sec:bs2pred}.

%% file: sections/predictions.tex
\section{Bias predictions versus simulations}
\label{sec:predictions}

In this section we confront the \whmpbs predictions with $N$-body measurements, step by step: the linear bias (Section~\ref{sec:b1}), the higher-order and tidal biases at fixed mass (Section~\ref{sec:bN_mass}), and the same biases at fixed $b_1$: the density relations and the selection prior they define (Section~\ref{sec:bN}), and the tidal sector (Section~\ref{sec:bs2pred}).
Before diving in, let us fix the rule by which every comparison is read. The \emph{mean} halo bias is the only statistic protected by closure: thanks to Eq.~\eqref{eq:envbias} it equals the direct halo response, no matter how deep the web hierarchy is taken --- the full nested sheet$\to$filament$\to$halo, the two-level filament$\to$halo, and the ``one-level'' halo evaluation all return the same value. The host \emph{median} enjoys no such protection: it is a non-linear functional of $P(b|\Mh)$ that shifts when a third web level is added, so we report it only as a descriptive statistic of the skew. 
All comparisons in this section are shown at $z=0$, but nothing hangs on that choice: every \whmpbs quantity is a function of the variance $S=\sigma^2(M,z)$ alone, so in $\nu=\delta_\mathrm{sc}/\sigma$, equivalently in $b_1$, the distributions and relations below are redshift-universal, and a different redshift merely relabels the mass axes.
The same argument holds not only for changes in redshift but also in \emph{cosmology}. Within the Markovian walk the power spectrum enters the first crossing only through $S=\sigma^2(M)$. Expressing every quantity against $b_1$ removes the entire dependence on the shape \emph{and} amplitude of $P(k)$, so the $b_N(b_1)$ relations are cosmology-universal.\footnote{This exact cancellation is a property of the Markovian approximation (Section~\ref{sec:walk}). A consistent non-Markovian treatment of the top-hat filter would make the first crossing depend on the local spectral slope $\mathrm{d}\ln\sigma^2/\mathrm{d}\ln M$ \citep{mussosheth2012,scs2013}, reintroducing a weak dependence on the \emph{shape} of $P(k)$ at fixed $b_1$. This does not affect redshift universality as long as changes in redshift are represented by a pure amplitude ($D^2(z)$) rescaling.} There is, however, another ingredient that can break redshift and cosmology universality: the critical overdensity threshold $\dc=\dc((\Omega_m(z),w(z)))$, which depends mildly on the matter density and dark energy equation of state evolution \citep{mead2017}. We leave investigating the impact of this on the priors derived here to future work.

\subsection{Linear bias: a distribution, not a number}
\label{sec:b1}
\begin{figure*}
\centering
\includegraphics[width=\textwidth]{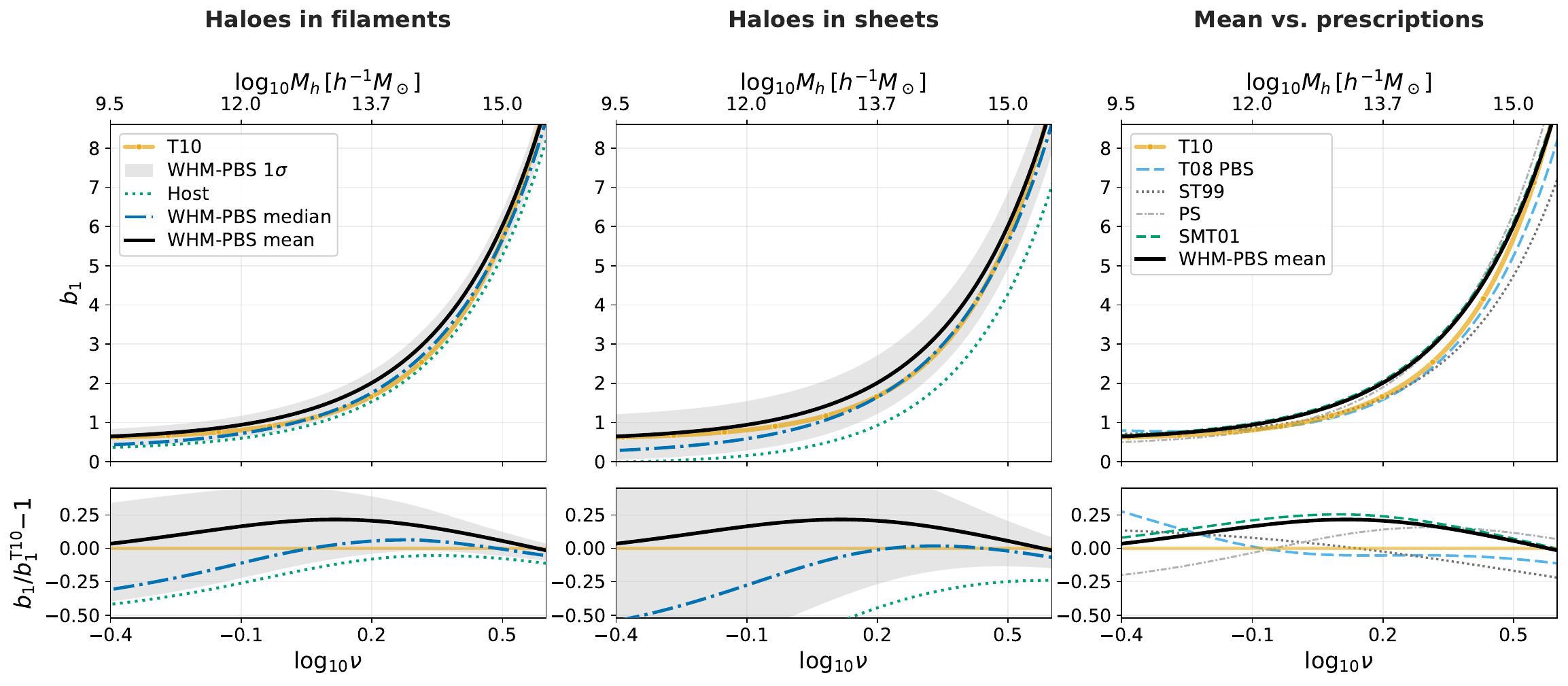}
\caption{The linear bias against peak height $\nu=\delta_\mathrm{sc}/\sigma$ (top axis: $\log_{10}$ halo mass), to compare with Fig.~1 of T10~\cite{tinker2010} and L16~\cite{lazeyras2016}. \emph{Left and middle:} the \whmpbs linear-bias distribution for the two-level (halo-in-filament) and three-level (halo-in-sheet) hierarchies, compared against the T10 calibration (orange). We show the mean (black solid; identical in both, since it is \emph{level-independent} by closure), the host median (blue dash-dotted), the \whmpbs $1\sigma$ host band (the 16--84 percentile, grey), and the host bias evaluated at the halo's own mass (green dotted: the $b_1$ a filament, left, or sheet, middle, of mass $M_h$ would have, a lower bound on the host bias since a halo's actual host is more massive). The median crosses the calibration toward high mass. \emph{Right:} the \whmpbs mean versus the state-of-the-art prescriptions: the T08 and ST99 PBS biases, Press--Schechter \citep[PS,][]{press1974} (spherical collapse), and the SMT01 ellipsoidal barrier evaluated with the same exact response. Each column has a lower panel showing the ratio to the T10 reference. The \whmpbs mean over-predicts by ${\sim}18\%$ and is \emph{not} a better linear-bias predictor than ST99 ($8\%$) or T08 ($5\%$). The novel \whmpbs content is the distribution and the $b_N(b_1)$ relations (Section~\ref{sec:bN}).}
\label{fig:b1}
\end{figure*}

The environment-averaged bias \eqref{eq:envbias} predicts, at fixed halo mass, not a single bias but the full distribution of biases that haloes of that mass inherit from their range of host
environments,
\begin{equation}
\label{eq:Pb1}
P(b_1|\Mh)=\int P(\Mf|\Mh)\,\delta_D\!\big(b_1-b_1(\Mf)\big)\,\mathrm d\Mf
\end{equation}
(mapped to Eulerian space, with the default \emph{filament} host of Section~\ref{sec:nested}; we quantify the broader sheet-host alternative below). Because $P(\Mf|\Mh)$ is broad and skewed
(Fig.~\ref{fig:cond}), so is $P(b_1|\Mh)$: it inherits a long tail to high bias from the rare, massive host filaments.\footnote{In fact, since $b_1(\Mf)$ is monotonic in the host mass, the fixed-$\Mh$ slices in the middle column of Fig.~\ref{fig:cond} [panels (e)/(k)] \emph{are} $P(b_1|\Mh)$, up to the relabelling $\Mf\to b_1(\Mf)$, and the right column [(f)/(l)] is the same distribution weighted by the bias it delivers.} The results are shown in Fig.~\ref{fig:b1}. Its two left columns present the \whmpbs distribution (the mean, the host median, and the $1\sigma$ host band) for the two-level (halo-in-filament) and three-level (halo-in-sheet) hierarchies, while its right column compares the \emph{mean} to other predictions in the literature.

Consider first the mean. By closure it is the standard large-scale bias, computed with the exact direct ($\delta$-response) bias \eqref{eq:pbsbias}. It captures the overall rise of $b_1$ with mass, but over-predicts the Tinker et al.~\citep[T10,][]{tinker2010} reference by $20\%$ at $\nu\!\simeq\!1$ and $7\%$ at $\nu\!\simeq\!3$. The T10 fit is itself confirmed to within $5\%$ by the separate-universe (and so exact peak--background--split) measurement of Lazeyras et al.~\citep[L16,][]{lazeyras2016}. We make no attempt to remedy this, and we do \emph{not} claim the mean as a better linear-bias predictor: it is not. Against the same benchmark the standard predictions do \emph{better}: the Sheth and Tormen~\citep[ST99,][]{shethtormen1999} and Tinker et al.~\citep[T08,][]{tinker2008} PBS biases lie within $8\%$ and $5\%$, and the simulation-calibrated excursion-set-peaks prediction \cite{Paranjape2013ESP} within ${\sim}2\%$. The over-prediction is the price we pay for a barrier calibrated to the cosmic-web morphology (sheets, filaments, haloes) rather than tuned to the bias. The same S06 barrier under-predicts the high-mass T08 mass function (Appendix~\ref{app:pbsbias}), the well-documented amplitude limitation of analytic ellipsoidal-collapse bias. The value of \whmpbs is therefore not the accuracy of this mean but the \emph{distribution} around it. In fact, every cosmological full-shape analysis marginalises the large-scale $b_1$ with a flat prior, so what the data actually constrain are the \emph{relative} relations $b_N(b_1)$ (Sections~\ref{sec:bN},~\ref{sec:assembly}), which are less sensitive to the \emph{absolute} $b_1(M)$ miscalibration. Every headline result of this paper depends on the former, not the latter.

The \emph{median} halo bias is the bias haloes inherit from their \emph{typical} host environment. Its mass $\Mf^{\rm med}(\Mh)$ is found via
\begin{equation}
\label{eq:median}
\int_{0}^{\,\Mf^{\rm med}(\Mh)} P(\Mf|\Mh)\,\mathrm{d}\Mf=\tfrac12 \qquad
b_1^{\rm med}(\Mh)=b_1\!\big(\Mf^{\rm med}(\Mh)\big) ,
\end{equation}
and by mapping towards the (monotonic) linear bias $b_1(\Mf)$ the median of the bias distribution $P(b_1|\Mh)$ of Eq.~\eqref{eq:Pb1} is found. It lies well below the mean, which is pulled up by the high-mass skew. Interestingly, $b_1^{\rm med}(\Mh)$ crosses the T10 reference toward the high-mass end (Fig.~\ref{fig:b1}, left and middle lower panels), so that at low mass T10 is bracketed by the median from below and the mean from above, while at high mass it runs along the median. We investigated whether this could be explained by any finite-box effects, but given the large boxes T10 used (up to $L_\mathrm{box}=1280\,\mathrm{Mpc}/h$) there is no such explanation. Hence, we flag this finding as a \emph{coincidence} (that also depends on the sheet versus filament only level) rather than a prediction and instead treat the median throughout as a descriptive statistic of the distribution's skew. 

The distribution's width, plotted as the grey $16$--$84$ band, is mass-dependent: it rises steadily with halo mass, so a single number is only indicative. Measured as the standard deviation $\sigma(b_1)=\sqrt{\mathrm{Var}(b_1|\Mh)}$, it is of order $\sigma\!\approx\!0.6-0.7$ for filament hosts and $\approx\!1.0$ for the broader nested filament-in-sheet host. Which level better matches the true scatter can only be settled with the help of N-body simulations. For example, Paranjape et al.~\cite{paranjape2018} find $\sigma_{b_1}\approx 3$ using their halo-by-halo bias estimator, but this is dominated by the sample noise depending on the adopted smoothing scale (appendix B1 therein) and cannot be compared to our prediction for \emph{intrinsic} scatter as function of host mass. On the other hand, we will see in Section~\ref{sec:pk}, the halo stochasticity measurement of Baldauf et al.~\cite{baldauf2013} favours the narrower filament host scheme.

To summarise the central qualitative \whmpbs predictions: halo bias at fixed mass is not a number but an intrinsically scattered, skewed distribution \citep{dekellahav1999,seljakwarren2004}, something no deterministic $b(M)$ can express. Its \emph{mean} is the closure-protected, level-independent value; its \emph{median} indicates the skewness and together with the \emph{mean} brackets the N-body measurements. Its \emph{width} is level-dependent (as the \emph{median}) and the narrower filament and broader sheet hosts encompass a plausible range for it, $\sigma(b_1)\!\approx\!0.6$--$1.0$. We show both levels throughout, and it is this range that later sets the widths of the \whm priors (Sections~\ref{sec:bN},~\ref{sec:pk}).

\subsection{Higher-order and tidal bias versus mass}
\label{sec:bN_mass}

\begin{figure*}
\centering
\includegraphics[width=\textwidth]{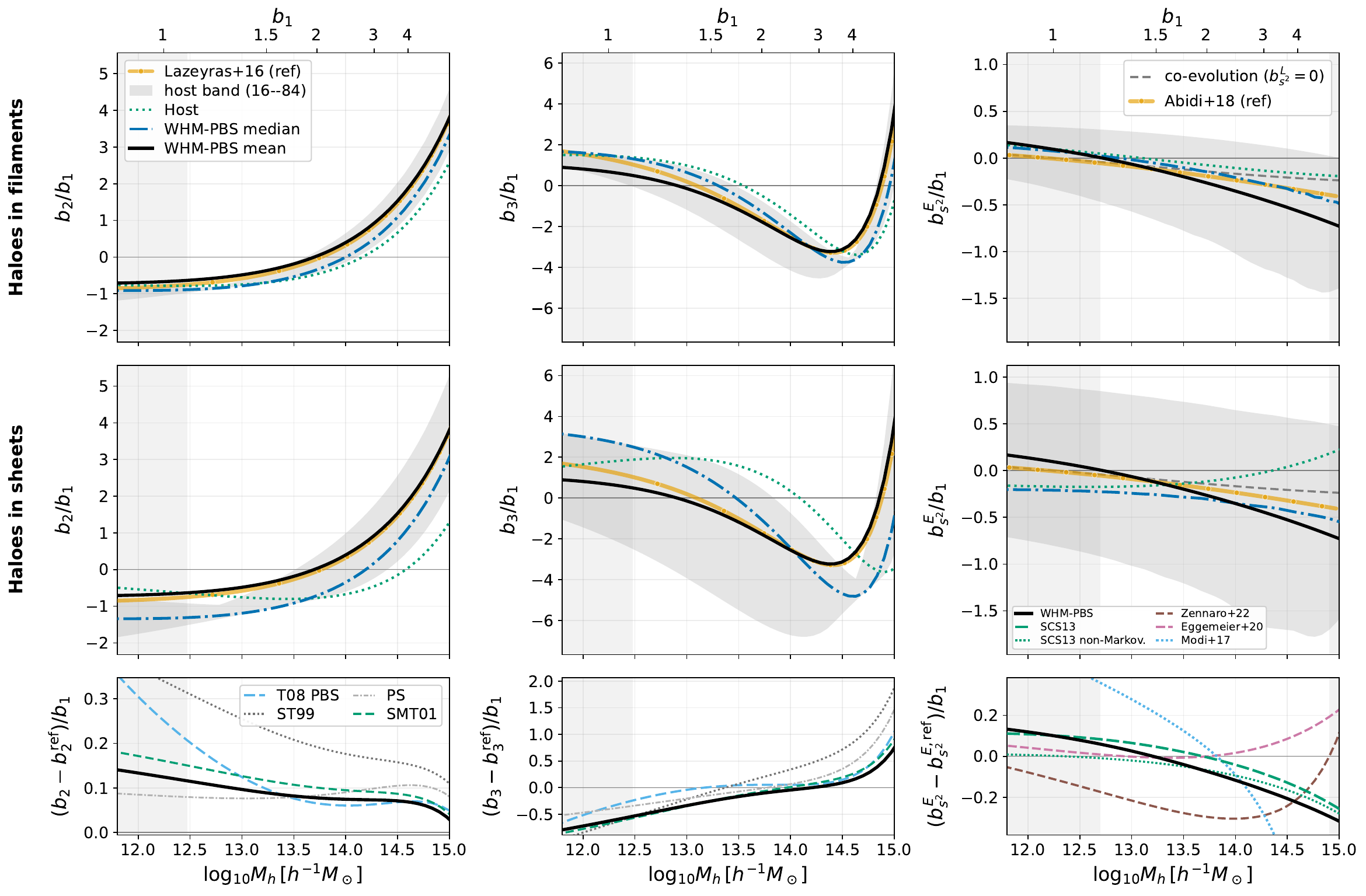}
\caption{The higher-order density biases $b_2,b_3$ and the Eulerian tidal bias $\bso^{E}=b_{\mathcal{G}_2}$ as functions of halo mass $M_h$ (top axis: the cosmology-robust $b_1$), in the same conventions as Figs.~\ref{fig:bNdens} and~\ref{fig:bNtid}, only against mass rather than $b_1$. \emph{Top two rows:} the distribution for haloes in filaments (narrower) and in sheets (broader), with the closure mean (black; $=$ direct halo, level-independent), the host median (blue dash-dot), the \whmpbs host band (grey: the range of $b_N$ spanned by the central $68\%$ of hosts, i.e.\ those between the $16$th and $84$th percentile in mass), and the unconditional host bias at the halo's own mass (green dotted); the measured references (orange) are L16~\cite{lazeyras2016} for $b_2,b_3$ and AB18~\cite{abidi2018} for $\bso^{E}$. \emph{Bottom row:} the residual of the mean against the reference, with the same comparison prescriptions as the $b_1(M_h)$ figure for $b_N$, plus the two SCS13~\cite{scs2013} excursion-set responses and the $N$-body fits of Zennaro et al.~\cite{zennaro2022}, Eggemeier et al.~\cite{eggemeier2020} and Modi et al.~\cite{modi2017} for $\bso^{E}$. The band measures the fixed-mass host scatter and by construction it always encloses the median host. Grey vertical shading marks extrapolation beyond the references' calibrated range.}
\label{fig:bNmass}
\end{figure*}

What holds for the linear bias holds at every order: at fixed mass the higher-order and tidal biases are distributions, not numbers. Figure~\ref{fig:bNmass} shows the predicted $b_2$, $b_3$, and the Eulerian tidal bias $\bso^{E}=b_{\mathcal{G}_2}$ against halo mass, each carrying the same $16$--$84$ host band as the linear bias, the assembly-bias amplitude inherited from the spread of host environments. The figure shows both host levels, the (default) narrower filament and the broader sheet, whose band over-predicts the amplitude (Section~\ref{sec:b1}); the residual row compares the mean against the measured references --- L16 for $b_2,b_3$ and Abidi and Baldauf~\cite[AB18,][]{abidi2018} for $\bso^{E}$ --- together with the same prescriptions as the $b_1$-figure \ref{fig:b1} and the $b_N(b_1)$ figures \ref{fig:bNdens} and \ref{fig:bNtid}, to which we defer for a more detailed comparison with the literature not repeated here.  

The figure shows two features worth commenting. 

First, both host relations $b_2(\Mf)$ and $b_3(\Mf)$ are non-monotonic: each has a minimum and turns over with mass, unlike the monotonic $b_1$. So the mapping $b_N(\Mf)$ is many-to-one rather than one-to-one (as for $b_1$). Hence, we built the $b_N$ bands and medians as follows: At each halo mass we rank the hosts by $b_1$ (equivalently by mass) and read the median $b_N$ off the $50$th-percentile host, while the grey band is the full range of $b_N$ spanned by the central $68\%$ of hosts (those between the $16$th and $84$th percentile in mass). Where a relation passes through its turnover, $b_N$ is stationary in mass, so a spread in host mass maps to only a small spread in $b_N$ and the band tightens. But it stays finite and always encompasses the median. It does not always encompass the mean, and we will come back to this in the next subsection.

Second, in the tidal column the \emph{filament} median hugs the co-evolution value ($\bso^{L}\!=\!0$, grey dashed) and the AB18 reference, whereas the sheet median departs from both. The filament barrier is nearly flat and so sources little tidal bias of its own, so a halo in a typical filament inherits close to pure co-evolution.

The next two sections re-express these biases against $b_1$ rather than mass, where the density and tidal scatter diverge: the density relations tighten to near-deterministic curves, while the tidal relation stays genuinely broad. We explain how to turn this predicted host scatter into priors for full-shape analyses.

\subsection{Density bias relations \texorpdfstring{$b_2(b_1),b_3(b_1)$}{b2(b1), b3(b1)}}
\label{sec:bN}
\begin{figure*}
\centering
\includegraphics[width=0.96\textwidth]{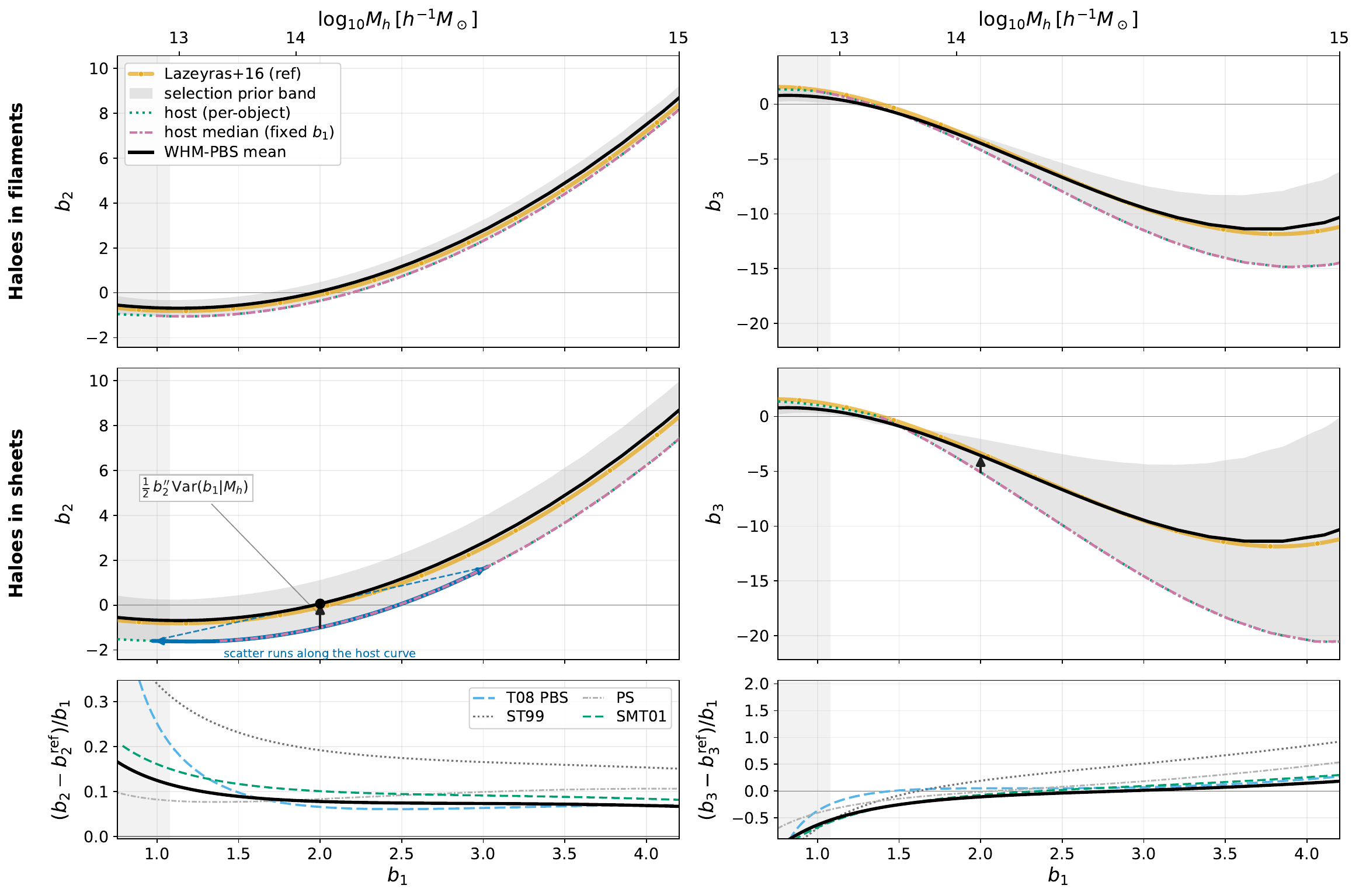}
\caption{Density biases $b_2$ (left) and $b_3$ (right) as \emph{relations} in $b_1$ from the moving-barrier response (top axis: $\log_{10}$ halo mass $M_h$). \emph{Top two rows:} the $b_N(b_1)$ relation for haloes in \emph{filaments} (the narrower default) and in \emph{sheets} (broader), showing the two curves of the aggregation rule \eqref{eq:aggregation}: the \emph{per-object host} relation (green dotted; the curve any single host-selected object obeys) and the \emph{mass-binned mean} (black; $=$ direct halo by closure, level-independent, identical in the two rows), which sits off the host curve by the sample's bias variance. The vertical arrow, drawn on the \emph{sheet} for better visualisation, marks this aggregation lift $\tfrac12 b_N''\,\mathrm{Var}(b_1|\Mh)$ in $b_N$ (equal to $\mathrm{Var}(b_1|\Mh)$ since $b_2''{=}2$), to scale on the raw-$b_N$ axis; it points up only, because the host curve is an exact lower envelope a sample cannot fall below. The shading is the resulting \emph{selection prior band}, Eq.~\eqref{eq:priorband}: centred on the mass-binned mean (black), with half-width $\mathrm{Var}(b_1|\Mh)$ (the aggregation lift) and the host curve as its hard floor $1\sigma$ below. The purple dash-dotted curve is the typical-host track of Fig.~\ref{fig:bNtid}, the fixed-mass median drawn at its own median $b_1$. Of course, for the density biases it collapses onto the per-object host curve itself. The blue arrow on the sheet row's $b_2$ panel visualises the mechanism: the fixed-mass scatter runs \emph{along} the host curve (drawn over $\pm1\sigma(b_1|\Mh)$ of this row's host distribution); the sample average (black dot) sits at the midpoint of the dashed chord joining the endpoints, lifted above the curve by exactly the variance. For the convex $b_2$ the host curve is an exact lower envelope; for the cubic $b_3$ the lift changes sign at low $b_1$ (the skewness term of Eq.~\eqref{eq:aggregation}). The sheet lift is always larger, its host distribution being wider than the filament host one. The measured relations of L16~\cite{lazeyras2016} (orange, the residual-row reference marked ``(ref)'') follow the mean closely, slightly pulled towards the median. \emph{Bottom row:} the residuals $(b_N-b_N^{\rm ref})/b_1$ with respect to the L16 reference. The \whmpbs mean (black) stays near zero, with the four density prescriptions of Fig.~\ref{fig:b1}. Vertical grey shading marks extrapolation beyond the calibrated L16 range $3\times10^{12}$--$10^{15}\,h^{-1}\Msun$.}
\label{fig:bNdens}
\end{figure*}

Having obtained all biases as a function of mass, we now change basis from $M_h$ to $b_1$. The predicted $b_2$ and $b_3$, again the \emph{mean} of the distribution, track the universal relations measured by L16 
with better precision than ST99, SMT01, PS, and similar precision as T08 for most of the $b_1$-range. 
It is crucial that these relations are expressed in $b_1$ rather than mass, since $b_1$ is the cosmology-robust variable: the ${\sim}18\%$ amplitude offset of the linear bias (Section~\ref{sec:b1}) is just a shift \emph{along} the $b_1$ axis, to which the barrier-\emph{shape}-driven $b_N(b_1)$ relations are mostly insensitive. And it is $b_N(b_1)$, not the absolute $b_1(M)$, that cosmological analyses actually use. Note also that here, unlike for the linear bias, it is the mean, the closure-protected and level-independent prediction, that follows the data. The median lies well below it, indicating the long high-mass tail of the host mass function making $P(b_N|\Mh)$ strongly skewed. That fixed-mass distribution, with its mean--median skew, was the subject of the versus-mass Fig.~\ref{fig:bNmass}. Figure~\ref{fig:bNdens} instead plots the $b_N(b_1)$ \emph{relations}. The scatter runs \emph{along} the haloes' host bias (per-object) relation; and what a given halo \emph{sample} adds on top of this per-object relation is fixed by an aggregation rule, as we now explain.

The step from this per-halo curve (determined by the host bias relation) to the measured halo bias relation is worth taking slowly. No survey or simulation deals with a `single halo': a mass bin, or a galaxy selection, is one \emph{composite} tracer whose effective bias parameters are the number-weighted means of its members. Averaging over a \emph{curved} relation does not commute with the curve, so a sample sits systematically \emph{above} the per-object $b_2(b_1)$ curve even though every member lies exactly on it. As a minimal illustration, take the convex $b_2(b_1)=b_1^2$: two haloes with $b_1=1$ and $3$ both lie exactly on the curve, yet the sample averages to $\langle b_2\rangle=5$, above the value $b_2(\langle b_1\rangle)=b_2(2)=4$ read off the curve at their mean.
Because the density biases are inherited from the host, averaging the per-object (host) relation $b_N^{\rm host}(b_1)$ over any sample $S$ and expanding about the sample's own linear bias $b_1^{S}=\langle b_1\rangle_S$ gives the \emph{aggregation rule}
\begin{equation}
\label{eq:aggregation}
b_N^{S}=\big\langle b_N^{\rm host}(b_1)\big\rangle_S
=b_N^{\rm host}\big(b_1^{S}\big)
+\tfrac12\,{b_N^{\rm host}}''\,\mathrm{Var}(b_1|S)
+\tfrac16\,{b_N^{\rm host}}'''\,\mu_3(b_1|S) + ...~.
\end{equation}
In this series expansion the first-derivative term vanishes ($\langle b_1-b_1^S\rangle_S=0$) by definition. Since $b_N(b_1)$ can be described by a polynomial of degree $N$ to very good accuracy (within the PS formalism ${b_2^{\rm host}}''=2$ and ${b_2^{\rm host}}'''=0$ exactly), only the terms up to third order are relevant here. But for $N>3$ higher moments need to be taken into account. Note that, in this case, the rule is \emph{exact} even though the host distribution is broad and skewed rather than Gaussian: the average of a degree-$N$ polynomial depends on only the first $N$ moments of $b_1$, whatever the distribution's shape.

This rule is sketched in figure \ref{fig:bNdens}. Since $b_2^{\rm host}$ is convex, the (host) per-object curve is an exact \emph{lower envelope} for $b_2$: every sample lies on or above it, lifted by its own bias variance. The mechanism is straight-forward: the fixed-mass scatter slides haloes \emph{along} the convex host $b_N(b_1)$ relation as indicated by the blue arrow in Fig.~\ref{fig:bNdens}, so averaging points spread along the curve lands the sample \emph{above} it, by the sagitta (the height of the host arc above the chord joining its $\pm\sigma$ endpoints), equal to $\tfrac12 b_2''\,\mathrm{Var}(b_1|S)$, which for the Press--Schechter (PS) curvature $b_2''{=}2$ is just $\mathrm{Var}(b_1|S)$ itself, as indicated by the vertical arrow of the middle left panel of Fig.~\ref{fig:bNdens}.
For the cubic $b_3^{\rm host}$ the curvature changes sign and the skewness term takes over at low $b_1$, so there the lift
changes sign with it, visible in the right-hand panels where the mean crosses the host curve. A \emph{mass-binned} halo sample carries the full host variance of Section~\ref{sec:bN_mass}, which for the default filament is itself mass-dependent, rising from $\mathrm{Var}(b_1|\Mh)\!\approx\!0.35$ at low mass to $\approx\!0.49$ at high mass, and is therefore lifted onto the mass-binned mean by that same amount ($\tfrac12 b_2''\,\mathrm{Var}=\mathrm{Var}$, again because $b_2''{=}2$), of order $+0.4$ but growing with mass. It is this curve, not the per-object relation, that the (mass-binned) separate-universe measurements of L16 test.

So far the aggregation lift has been, given a sample, a computable \emph{number}: Eq.~\eqref{eq:aggregation} says exactly how far above the per-object curve the sample sits. It becomes a \emph{prior} only once we admit that a real sample's $\mathrm{Var}(b_1|S)$ is not known in advance.
Equation~\eqref{eq:aggregation} is thus the prior recipe for a full-shape analysis, mapping onto the relation plane as the shaded selection prior band of Fig.~\ref{fig:bNdens} at the sample's measured $b_1$,
\begin{equation}
\label{eq:priorband}
b_2\,\big|\,b_1\;\sim\;\mathcal N\Big(b_2^{\rm mean}(b_1),\,
\big[\mathrm{Var}(b_1|\Mh)\big]^2\Big)
\quad\text{truncated at}\quad b_2\ge b_2^{\rm host}(b_1),
\end{equation}
with $b_2^{\rm mean}(b_1)=b_2^{\rm host}(b_1)+\mathrm{Var}(b_1|\Mh)$ the mass-binned mean relation [$\mathrm{Var}(b_1|\Mh)\!\approx\!0.35$--$0.49$, mass-dependent] and the width $\sigma_{b_2}=\mathrm{Var}(b_1|\Mh)$, so the host floor sits exactly $1\sigma$ below the centre. As shown in the Figure this prior encompasses the L16 reference.
We demonstrate this prior use in Section~\ref{sec:demo}.

\subsection{Tidal bias relations \texorpdfstring{$\bso(b_1),\bGa(b_1)$}{bs2(b1), bGamma3(b1)}}
\label{sec:bs2pred}
\begin{figure*}
\centering
\includegraphics[width=0.96\textwidth]{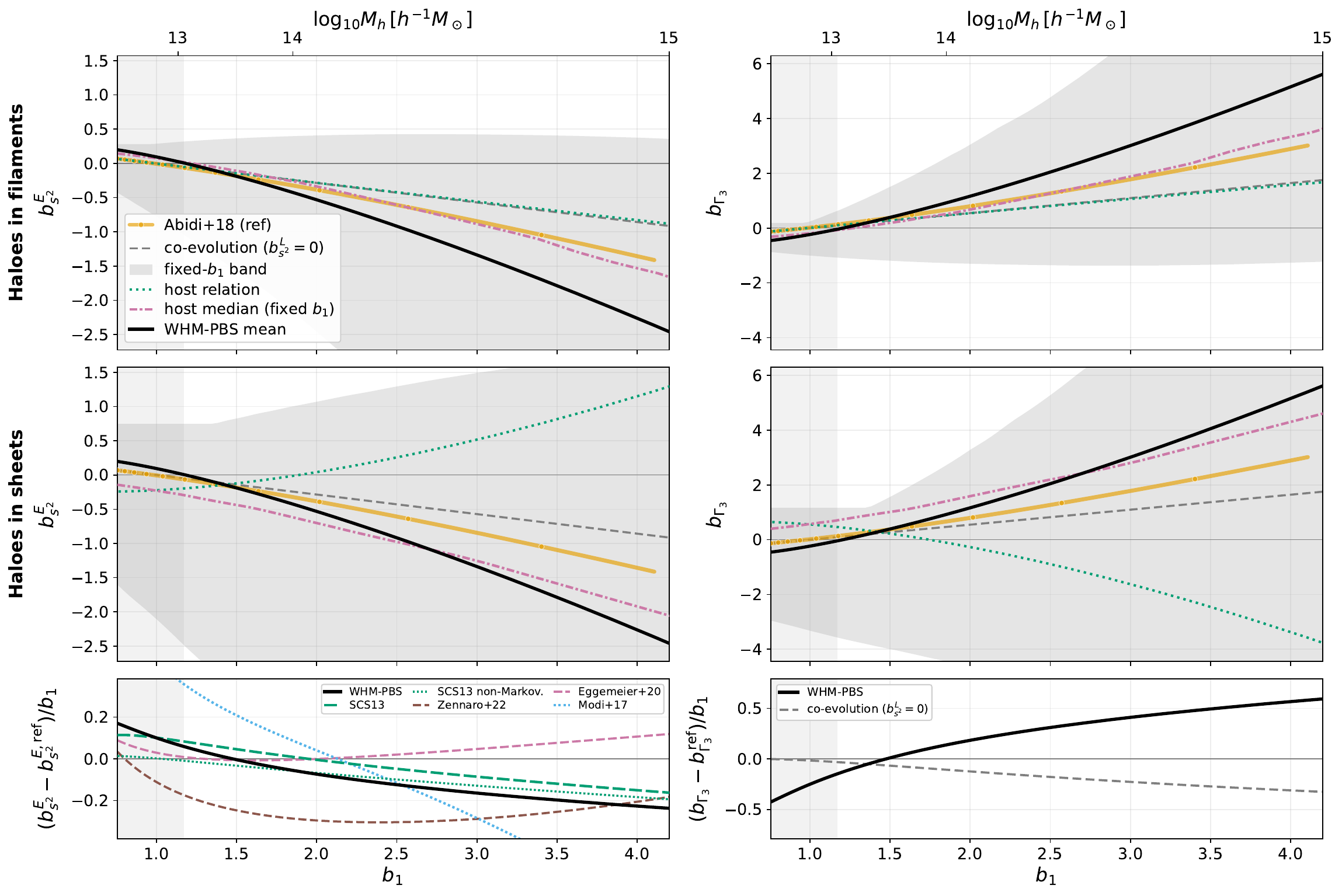}
\caption{The tidal sector in the Eulerian frame: $\bso^{E}=b_{\mathcal{G}_2}$ (left) and $\bGa$ (right) as functions of $b_1$ (top axis: $\log_{10}M_h$), in the three-row filament/sheet/residual format of Fig.~\ref{fig:bNdens}. Black: the closure mean; grey band: the fixed-$b_1$ per-object scatter of the broken relation (Section~\ref{sec:bs2pred}), clipped at the panel edges; green dotted: the host relation; purple dash-dotted: the typical-host track (the fixed-mass median drawn at its own median $b_1$); grey dashed: pure co-evolution ($\bso^{L}=0$, left; $\tfrac{23}{42}(b_1-1)$, right). Orange: the AB18~\cite{abidi2018} reference; the residual panels compare the mean, the SCS13~\cite{scs2013} predictions, and the fits of Zennaro et al.~\cite{zennaro2022}, Eggemeier et al.~\cite{eggemeier2020} and Modi et al.~\cite{modi2017} against the reference. The right panel is derived from $\bso^{L}$ through Eq.~\eqref{eq:bgamma3} with no new parameter. Grey vertical shading marks extrapolation beyond the AB18 calibrated range $5\times10^{12}$--$8\times10^{14}\,h^{-1}\Msun$.}
\label{fig:bNtid}
\end{figure*}

The exact tidal response \eqref{eq:tid-uncond} predicts the Lagrangian tidal bias $\bso^{L}$; the left panel of Fig.~\ref{fig:bNtid} shows it in the Eulerian frame $\bso^{E}=\bso^{L}-\tfrac27(b_1-1)$, equal to the Galileon bias $b_{\mathcal{G}_2}$ of the EFT basis, so that it shares the frame of the Eulerian $\bGa$. As the co-evolution shift is deterministic in $b_1$ it cancels in differences, so the residual panel and the comparisons below are frame-independent.
That residual panel compares the prediction against the AB18~\cite{abidi2018} reference, together with the $N$-body fits of Zennaro et al.~\cite{zennaro2022}, the Eggemeier et al.~\cite{eggemeier2020} fit to the Lazeyras and Schmidt~\cite{lazeyras2018} data, Modi et al.~\cite{modi2017}, and \emph{both} excursion-set predictions of Sheth et al.~\citep[SCS13,][]{scs2013} (their Markovian eq.~30 and the non-Markovian eq.~34).

Regarding its \emph{mean}, the tidal bias behaves much as the linear bias did. It steepens at high mass and over-predicts the AB18~\cite{abidi2018} amplitude by ${\sim}1.5\times$ at $b_1\!\simeq\!3$ in the Eulerian frame shown (${\sim}2.5\times$ in the Lagrangian, where the shared co-evolution offset no longer dominates). This is qualitatively the same over-prediction as for the linear bias (Section~\ref{sec:b1}), though larger in magnitude; both trace back to the same web-calibrated ellipsoidal barrier.

The grey band in Fig.~\ref{fig:bNtid} is not this mean but the genuine fixed-$b_1$ scatter of the \emph{broken} $\bso(b_1)$ relation, and it brackets the full spread of the $N$-body fits, from the shallow AB18 to the steep Zennaro et al.~\cite{zennaro2022}. Why is this relation broken, when the density relations of Section~\ref{sec:bN} are not?
This is best understood by considering Eq.~\eqref{eq:tid-envbias}. Both (density and tidal) sectors average over the \emph{same} host distribution $P(\Mf|\Mh)$; they differ only in the weight inside the integral. For the density biases the weight is the host's own bias $b_{f,N}(\Mf)$ [Eq.~\eqref{eq:envbias}], a function of the host mass alone: $b_1$ and $b_N$ are two read-outs of the single random variable $\Mf$, so fixing one fixes the other, and the host scatter can only slide objects \emph{along} the relation. For the tidal bias the weight is the \emph{conditional} response $\bso^{L}(\Mh|\Mf)$, a function of the (halo, host) \emph{pair} through the barrier stretch between them, while the halo's $b_1$ is still inherited from the host alone. Fixing $b_1$ therefore fixes only the host: every halo mass that the host can contain still carries its own conditional response, and the single tight curve opens into a genuinely two-dimensional distribution. 

Considering the other \whmpbs predictions, the median is more encouraging than the mean here. Following the \emph{typical} host -- at each mass, the median tidal response taken at the median inherited $b_1$ -- gives a prediction that follows the measured relation.
We interpret this with the same caution as the $b_1$ median -- it is not closure-protected, and the agreement is specific to our default filament host, the sheet median missing by $0.41$.

The two sectors reveal an instructive contrast. For the \emph{density} biases the same typical-host construction reproduces the per-object \emph{relation}, not the measurements: it rides the host curve (the purple track of Fig.~\ref{fig:bNdens}), below the mass-binned data. For the \emph{broken} tidal relation it is the other way around --- there the typical host is the better predictor of the measurements themselves. One might speculate that in the broken relation the gap between mean and median is set even more strongly by the rare, high-mass hosts, precisely the tail that finite-volume simulations sample poorly, so that the measurements sit nearer the typical host than the mean; but this is only a conjecture.

A curiosity of the comparison is worth mentioning. Although our walk is Markovian (sharp-$k$, uncorrelated steps), its \emph{mean} tracks the SCS13 \emph{non-Markovian} (correlated-step) prediction slightly more closely than their Markovian one. We take this as a coincidence rather than a correspondence, given that SCS13 adopt a very different moving barrier shape.

The same response predicts the cubic non-local bias $\bGa$. By the co-evolution relation \eqref{eq:bgamma3} of Section~\ref{sec:bgamma3}, the moving-barrier tidal bias $\bso^{L}$ fixes $\bGa=\tfrac{23}{42}(b_1-1)-\tfrac52\bso^{L}$ with no new free parameter. The result (Fig.~\ref{fig:bNtid}, right) sits systematically \emph{above} the pure co-evolution value $\tfrac{23}{42}(b_1-1)$ usually adopted as the EFT prior, lifted by the tidal bias, in the direction that the direct measurements of AB18~\cite{abidi2018}, Lazeyras and Schmidt~\cite{lazeyras2018}, and Eggemeier et al.~\cite{eggemeier2021} prefer over pure co-evolution. Being directly determined by $\bso^{L}$, $\bGa$ mirrors the conclusions obtained for the tidal bias: the closure mean overpredicts the measured amplitude by ${\sim}1.5\times$ at high mass, while the host band brackets it.

We now turn to establishing the tidal bias prior for the demonstration of Section~\ref{sec:demo}. As we shall see, the tidal bias prior has two ingredients: the first coming from the uncertainty on its \emph{variance} (the same selection prior as in the density bias cases), the second from the assembly-bias-induced uncertainty on the host's \emph{mean} itself. 

The first ingredient mirrors the selection prior from Section~\ref{sec:bN}, i.e., a curved relation turns the halo-variance along the host relation into a lift off that relation --- the host--mean band of Section~\ref{sec:bN}, $|b_N^{\rm mean}-b_N^{\rm host}|$. In exact analogy to Eq.~\eqref{eq:priorband}, we take here the host--mean band of $|\bso^{\rm mean}-\bso^{\rm host}|$ as the resulting prior band. 

The tidal relation is \emph{broken}, and so carries a second ingredient. A background shear is not absorbed by the host alone, so individual (halo, host) pairs scatter at fixed $b_1$ --- the grey band $\sigma_{\rm band}$ of Fig.~\ref{fig:bNtid}. A selection \emph{blind} to the host at fixed mass averages that scatter away and returns to the mean by closure, so the per-object band is not itself the prior; but a selection \emph{correlated} with the host shifts the sample's \emph{mean} off the relation, by some \emph{fraction} $\lambda_S$ of $\sigma_{\rm band}$. That fraction is the assembly-bias signal of the tidal sector that can be calibrated against simulations, for example the concentration- and spin-split measurements of Lazeyras et al.~\cite{lazeyrasbarreira2021}. We describe this procedure in Section~\ref{sec:tidal_assembly}, where we find that ranking haloes on an internal property reaches a ceiling of $0.35\,\sigma_{\rm band}$, i.e., $\lambda_S = 0.35$.\footnote{Another way to see that only a \emph{fraction} of the band should be used, is to consider the extreme case of $\lambda_S=1$. This basically means that some internal halo property entering the selection is 100\% correlated with the cosmic web environment, which is not observed in simulations.}

Adopting that ceiling as the conservative width, the two ingredients add in quadrature,
\begin{equation}
\label{eq:tid-priorband}
\bso\,\big|\,b_1\;\sim\;\mathcal N\Big(\bso^{\rm mean}(b_1),\;
\big[\bso^{\rm mean}-\bso^{\rm host}\big]^{2}
+\big[0.35\,\sigma_{\rm band}\big]^{2}\Big).
\end{equation}
The first term is the variance band shared with the density prior \eqref{eq:priorband}; the second is the mean shift the density sector does not need, sliding only along its relation. With the filament host as default, the band runs a factor $4$ tighter than the raw per-object one ($\lambda_S=1$) and holds the AB18~\cite{abidi2018} reference throughout its calibrated range. The cubic bias $\bGa$ is determined from $(b_1,\bso^{L})$ via Eq.~\eqref{eq:bgamma3}. Section~\ref{sec:demo} puts these priors to work, and Section~\ref{sec:conclusions} states them as a short set of rules.

\newpage

%% file: sections/stochasticity.tex
\section{Stochasticity from the bias distribution}
\label{sec:pk}
In Section~\ref{sec:bN} we used the width of $P(b_N|\Mh)$ to construct the prior band~\eqref{eq:priorband}, hence treating it as an \emph{uncertainty}. Taken as a \emph{random} field, it sources a stochastic, shot-noise-like contribution to the halo power spectrum, and this is what we develop in this section. Concretely, we develop the formalism to predict the full stochasticity \emph{matrix} across mass samples. 

Stochasticity is the part of a halo field that does not correlate with the matter. For a single mass sample, the standard formulation is to split the halo overdensity into a linear-bias piece and a residual (noise) piece,
\begin{equation}
\delta_h (\mathbf{x}) =b_1\,\delta (\mathbf{x}) +\varepsilon (\mathbf{x})~.
\label{eq:deltah_split}
\end{equation}
In Fourier space, one can form the halo auto- and cross-spectra $P_{hh}(k)=\langle\delta_h (\mathbf{k}) \delta_h (\mathbf{k}) \rangle$ and $P_{hm}(k)=\langle\delta_h (\mathbf{k}) \delta (\mathbf{k})\rangle$. Inserting the Fourier transform of equation \eqref{eq:deltah_split} and suppressing the $(k)$ argument, gives
\begin{equation}
P_{hm}(k) =b_1\,P_{mm}+\langle\varepsilon\delta\rangle,\qquad
P_{hh}(k) =b_1^2\,P_{mm}+2b_1\langle\varepsilon\delta\rangle+\langle\varepsilon\varepsilon\rangle~.
\label{eq:PhhPhm}
\end{equation}
Since the noise is uncorrelated with the matter, $\langle\varepsilon\delta\rangle=0$, such that $P_{hm}=b_1P_{mm}$ and the stochastic power is the excess of the halo auto spectrum over its matter-correlated part,
\begin{equation}
P_\varepsilon\equiv \langle\varepsilon\varepsilon\rangle = P_{hh}-\frac{P_{hm}^2}{P_{mm}}~.
\label{eq:Peps_def}
\end{equation}
Across halo mass bins $i,j$ with halo overdensity fields $\delta_{h,i}, \delta_{h,j}$, and linear halo bias $b_i, b_j$, respectively, this generalises to the full stochasticity matrix,
\begin{equation}
\sigma_{ij}(k)=\big\langle(\delta_{h,i}-b_{1,i}\delta)(\delta_{h,j}-b_{1,j}\delta)\big\rangle,
\label{eq:Cij}
\end{equation}
whose diagonal is the single-sample residual power $P_\varepsilon$ of sample $i$. 
It consists of three types of ingredients:
\begin{itemize}
\item \emph{Poisson contribution:} This is the standard ``shot-noise'' term corresponding to the inverse number density $1/n_i$ of bin $i$ and only appears in the diagonal.
\item \emph{Super-Poisson contribution}: This is mostly referred to as the ``clustering'' contribution: haloes cluster together in shared overdense environments beyond the linear-bias expectation. It is a non-linear bias effect: by the definition~\eqref{eq:Cij} a quadratic response $\propto b_2\delta^2$ enters as noise rather than being absorbed into $b_1$ (see \cite{mcdonaldroy2009} for the renormalisation of shot noise by non-linear bias). As we will motivate in Section~\ref{sec:superP}, \whmpbs computes it nonperturbatively as the within-host bias scatter of $P(b_N|\Mh)$.
\item \emph{Sub-Poisson contribution}: This (negative) contribution is due to the effect of halo exclusion, for which we take the model of Baldauf et al.~\cite[B13,][]{baldauf2013}, briefly summarised in Section~\ref{sec:subP}. It also impacts all stochasticity matrix elements.
\end{itemize}
Written as the sum of these three ingredients, the stochasticity matrix reads
\begin{equation}
\sigma_{ij}(k)=\frac{\delta_{ij}}{\bar n_i}+P^{\rm clust}_{\varepsilon,ij}(k)+P^{\rm excl}_{\varepsilon,ij}(k)~,
\label{eq:stochsum}
\end{equation}
with the Poisson floor $\delta_{ij}/\bar n_i$ on the diagonal alone, and the super-Poisson clustering $P^{\rm clust}_{\varepsilon,ij}$ and sub-Poisson exclusion $P^{\rm excl}_{\varepsilon,ij}$ developed in turn below. 
We compare our parameter-free predictions to the model and data from B13 in Section~\ref{sec:pkcomp}.

\subsection{Super-Poisson stochasticity (clustering)} \label{sec:superP}
The key step in \whmpbs is that the bias is no longer a single number but a \emph{field}: a halo inherits the bias of whatever host it occupies, and that host environment (taken here as a filament) changes from place to place, so the linear halo bias is position dependent. We make this dependence explicit and write
\begin{equation}
\begin{aligned}
b_1(\bm x|\Mh)&=b_1(\Mf(\bm x|\Mh)) \\
&= \langle b_1(\Mf(\bm x|\Mh)) \rangle +\delta b(\bm x) \\
&=\langle b_1\rangle_{\Mf|\Mh}+\delta b(\bm x)~,
\end{aligned}
\label{eq:biasfield}
\end{equation}
where $\Mf(\bm x|\Mh)$ is the host-mass field, the mass of the host occupied by the halo of mass $\Mh$ at $\bm x$, and $b_1(\Mf)$ the universal host-bias relation, so the local bias is position dependent solely through $\Mf(\bm x)$; where unambiguous we drop the conditioning on $\Mh$ to lighten the notation. Two averages enter in \eqref{eq:biasfield}, and we keep them typographically distinct: the \emph{host average} $\langle\,\cdot\,\rangle_{\Mf|\Mh}\equiv\int\!\mathrm d\Mf\,P(\Mf|\Mh)\,(\,\cdot\,)$ of the environment-averaged bias \eqref{eq:envbias}, and the \emph{ensemble average} $\langle\,\cdot\,\rangle$ over realisations of the density field, which defines the power spectra we compute below. The second line of \eqref{eq:biasfield} decomposes the bias field into its ensemble mean and the offset $\delta b(\bm x)$ about it; 
the third follows from the \emph{assumption} that the ensemble average of any function of the host mass is its host average, $\langle f(\Mf(\bm x))\rangle=\langle f\rangle_{\Mf|\Mh}$.\footnote{For a general random field, there is no reason to believe that both these types of averages would be equivalent. In our case, the imposed field-nature of linear bias is ``artificial'': within the \whmpbs it is still deterministic at the level of the host, i.e., it is purely the host mass $\Mf$ that determines the positional dependence here.}

Applying this bias field to the matter splits the halo overdensity \eqref{eq:deltah_split} into a deterministic and a fluctuating piece,
\begin{equation}
\delta_h=b_1(\bm x)\,\delta
=\langle b_1\rangle_{\Mf|\Mh}\,\delta+\underbrace{\delta b(\bm x)\,\delta}_{\displaystyle\varepsilon^{\rm clust}}~,
\label{eq:deltah_field}
\end{equation}
so the residual left by this continuous bias field is the bias fluctuation acting on the matter, $\varepsilon^{\rm clust}=\delta b\,\delta$, the super-Poisson (clustering) part of the full residual $\varepsilon$ of \eqref{eq:deltah_split}.
Since the cross-spectrum is \emph{linear} in the bias,  only the mean survives:
\begin{equation}
\label{eq:Phmclosure}
\begin{aligned}
P_{hm}=\langle\delta_h\,\delta\rangle
&=\big\langle\,b_1(\bm x)\,\delta\;\delta\,\big\rangle
=\big\langle\big[\langle b_1\rangle_{\Mf|\Mh}\,\delta+\varepsilon^{\rm clust}\big]\,\delta\big\rangle\\
&=\langle b_1\rangle_{\Mf|\Mh}\,P_{mm}+\underbrace{\langle\varepsilon^{\rm clust}\,\delta\rangle}_{=\,0} \,.
\end{aligned}
\end{equation}
Here, the noise--matter term $\langle\varepsilon^{\rm clust}\delta\rangle$ vanishes, because the cross spectrum is the ensemble average of a quantity that depends on position solely via the host mass. \footnote{An alternative explanation is that the offset $\delta b$ is to leading order proportional to the matter field ($\delta b \propto b_2 \delta$, see Eq.~\eqref{eq:biasexp}), so $\varepsilon^{\rm clust}$ is a product of two density fields and its correlation with $\delta$, $\langle\varepsilon^{\rm clust}\delta\rangle=\langle\delta b\,\delta\,\delta\rangle$, is a three-point (odd) moment of the near-Gaussian matter field that vanishes at leading order.} Hence, the cross-spectrum measures the mean bias alone, blind to the scatter $P(b_1|\Mh)$, and the measured bias is unshifted.

The auto-spectrum is \emph{quadratic}, and there the same fluctuation cannot cancel:
\begin{equation}
\label{eq:Phhauto}
\begin{aligned}
P_{hh}=\langle\delta_h\,\delta_h\rangle
&=\big\langle\,b_1(\bm x)\,\delta\;b_1(\bm x)\,\delta\,\big\rangle
=\big\langle\big[\langle b_1\rangle_{\Mf|\Mh}\,\delta+\varepsilon^{\rm clust}\big]^2\big\rangle\\
&=\langle b_1\rangle_{\Mf|\Mh}^2\,P_{mm}
+2\langle b_1\rangle_{\Mf|\Mh}\underbrace{\langle\varepsilon^{\rm clust}\,\delta\rangle}_{=\,0}
+\underbrace{\langle\varepsilon^{\rm clust}\,\varepsilon^{\rm clust}\rangle}_{\displaystyle P_\varepsilon^{\rm clust}}\\
&=\langle b_1\rangle_{\Mf|\Mh}^2\,P_{mm}+P_\varepsilon^{\rm clust}\,.
\end{aligned}
\end{equation}
The cross term vanishes by the same argument, but the \emph{four-point} (even) moment $P_\varepsilon^{\rm clust}\equiv\langle\varepsilon^{\rm clust}\varepsilon^{\rm clust}\rangle$ does not: it is the excess of the halo auto over its matter-correlated part, $P_\varepsilon^{\rm clust}=P_{hh}-P_{hm}^2/P_{mm}$, precisely the residual power $P_\varepsilon$ of \eqref{eq:Peps_def}. To a large-scale observer that excess power is sourced by host-scale modes too small to resolve [the convolution \eqref{eq:Peps} below] and is indistinguishable from an added shot noise. This is the super-Poisson stochasticity. Computing it requires the statistics of the offset field $\delta b$, to which we now turn.

\whmpbs fixes  the clustering residual $\varepsilon^{\rm clust}$ from the same conditional host distribution that sets the mean. The offset $\delta b$ is not white noise: each host imprints $b_1(\Mf)-\langle b_1\rangle_{\Mf|\Mh}$ coherently across its \emph{own} Lagrangian scale $R_\star(\Mf)=(3\Mf/4\pi\bar\rho)^{1/3}$, so two haloes that share a host carry the same offset while haloes in different hosts are independent. The bias-fluctuation power is therefore the squared offset, weighted by the host window and its volume, averaged over the host distribution,
\begin{equation}
\begin{aligned}
\label{eq:Pdb}
P_{\delta b}(q)&=\int\!\mathrm d\Mf\,P(\Mf|\Mh)\,\big[b_1(\Mf)-\langle b_1\rangle_{\Mf|\Mh}\big]^2\,
V_\star(\Mf)\,W_{R_\star}^2(q)~ \\
&\mathrm{with} \quad \int\!\frac{\mathrm d^3q}{(2\pi)^3}\,P_{\delta b}=\mathrm{Var}(b_1),
\end{aligned}
\end{equation}
where $W_{R_\star}(q) = W\!\big(q; R_\star(\Mf)\big)$ is the host-coherence window (a spherical top-hat equal to one for $q\!\ll\!1/R_\star$ and cutting off for $q\!\gtrsim\!1/R_\star$) and $V_\star(\Mf)=\tfrac{4\pi}{3}R_\star^3$ the host volume.

Because $\varepsilon^{\rm clust}$ is the \emph{product} of two (real space) fields, the bias fluctuation and the matter, its (Fourier space) power spectrum is their \emph{convolution},
\begin{equation}
\label{eq:Peps}
P_\varepsilon^{\rm clust}(k)=\int\!\frac{\mathrm d^3q}{(2\pi)^3}\,P_{\delta b}(q)\,P_{mm}(|\bm k-\bm q|).
\end{equation}
This single integral carries all the scale dependence of the stochasticity, and its shape can be understood by considering the following two limits: For $k\!\ll\!1/R_\star$, most of the weight $P_{\delta b}(q)$ sits at $q\!\sim\!1/R_\star\!\gg\!k$, so displacing $\bm k$ barely moves $P_{mm}(|\bm k-\bm q|)\simeq P_{mm}(q)$ and the convolution freezes at a white plateau,
\begin{equation}
\label{eq:Pepsplateau}
P_\varepsilon^{\rm clust}(k\rightarrow 0) \longrightarrow \int \!\mathrm d\Mf\,P(\Mf|\Mh)\,\big[b_1(\Mf)-\langle b_1\rangle_{\Mf|\Mh}\big]^2\,V_\star(\Mf)\,\sigma^2(\Mf)~,
\end{equation}
with $\sigma^2(\Mf)=\int\!\mathrm d^3q/(2\pi)^3\,W_{R_\star}^2(q)\,P_{mm}(q)$ the matter variance on the host scale. This is the large-scale, super-Poisson, shot-noise-like excess. At smaller scales than the host scale, for $k\!\gg\!1/R_\star$, the support of $P_{\delta b}$ lies entirely at $q\!\ll\!k$, so $P_{mm}(|\bm k-\bm q|)\!\simeq\!P_{mm}(k)$ factors out of \eqref{eq:Peps} and the normalisation of $P_{\delta b}$ in \eqref{eq:Pdb} yields
\begin{equation}
\label{eq:Pepshighk}
P_\varepsilon^{\rm clust}(k\rightarrow \infty) \longrightarrow \mathrm{Var}(b_1)\,P_{mm}(k) \longrightarrow 0~.
\end{equation}
Since $P_{mm}(k)$ falls steeply, the stochasticity drops back towards the Poisson floor,
$\sigma_{ii}=1/\bar n+P_\varepsilon^{\rm clust}\!\to\!1/\bar n$, with the crossover at the host scale
$R_\star$.

Let's recapitulate and comment on the consequences of these findings. Below the host scale, $\delta b$ is effectively frozen, the auto-spectrum is merely boosted, $P_{hh}\!\to\!(\langle b_1\rangle_{\Mf|\Mh}^2+\mathrm{Var}(b_1))\,P_{mm}$, and the excess $\mathrm{Var}(b_1)P_{mm}$ is degenerate with a slightly larger linear bias $\sqrt{\langle b_1^2\rangle_{\Mf|\Mh}}$ and reabsorbed. Only \emph{above} the host scale, where the finite coherence of $\delta b$ makes it act as white noise, the variance becomes genuine stochasticity. Were the host infinitely large ($R_\star\!\to\!\infty$, a single bias per realisation), $P_{\delta b}$ would collapse to a delta function and $P_\varepsilon^{\rm clust}=\mathrm{Var}(b_1)P_{mm}$ at \emph{all} $k$, a pure bias renormalisation with no stochasticity at all. It is the finite host scale that turns the bias variance into an observable, white shot-noise excess. The plateau is accordingly set by the typical hosts, while the most massive hosts (largest $R_\star$, smallest $1/R_\star$) contribute only the reabsorbed, $P_{mm}$-shaped term.

It is worth pausing to think what this excess actually \emph{is}. By the peak--background split the environment modulates the linear bias, $b_1^{\rm eff}=\langle b_1\rangle_{\Mf|\Mh}+b_2\,\delta_{\rm env}$, so the host scatter is $\delta b= b_2\,\delta_{\rm env}$ at leading order, with $\mathrm{Var}(b_1)=b_2^2\,\mathrm{Var}(\delta_{\rm env})$, and the convolution \eqref{eq:Peps} is the \whmpbs version of the $\tfrac12 b_2^2\!\int\!\mathrm d^3r\,\xi^2$ nonlinear-clustering term of B13. The host scatter is therefore a clustering effect, not a new kind of noise: a denser environment raises the bias, so the halo field $\delta_h=b_1(\bm x)\,\delta$ clusters more strongly than $\langle b_1\rangle_{\Mf|\Mh}\,\delta$ exactly where $\delta$ is large, and the definition \eqref{eq:Peps_def} books this deterministic excess as noise because it does not correlate \emph{linearly} with the matter. In this sense it is \emph{effective} stochasticity: it is the only ingredient of \eqref{eq:stochsum} that does not rely on the discreteness of haloes --- it would survive for a continuous biased tracer, and it would vanish in a homogeneous universe, unlike the Poisson floor and the exclusion term. Resummed over the full host distribution rather than truncated at $b_2$, it is what $P(\Mf|\Mh)$ supplies with no parameter to fit.

Nothing in this is special to a single halo sample, and the same construction gives the whole stochasticity matrix $\sigma_{ij}$. All mass bins draw their offsets from the \emph{one} host-bias
function $b_1(\Mf)$: a bin-$i$ halo carries $\delta b_i(\bm x)=b_1(\Mf(\bm x))-\langle b_1\rangle_i$,
one universal function of the local host mass minus a bin-dependent constant, so haloes of different
mass are correlated wherever they share a host (haloes in different hosts correlate only through the matter, which Eq.~\eqref{eq:Cij} subtracts). The bias-fluctuation power \eqref{eq:Pdb} then
becomes a cross-power with a shared-host kernel in place of the single host distribution,
\begin{equation}
\label{eq:Pdbij}
P_{\delta b,ij}(q)=\int\!\mathrm d\Mf\,Q_{ij}(\Mf)\,\big[b_1(\Mf)-\langle b_1\rangle_i\big]
\big[b_1(\Mf)-\langle b_1\rangle_j\big]\, V_\star(\Mf)\,W_{R_\star}^2(q)~,
\end{equation}
with the shared-host kernel $Q_{ij}(\Mf)=\sqrt{P(\Mf|M_i)\,P(\Mf|M_j)}$, the geometric mean of the
two bins' host distributions: symmetric, supported only where \emph{both} bins find hosts, and
reducing exactly to \eqref{eq:Pdb} on the diagonal $i=j$.  The convolution \eqref{eq:Peps} carries $P_{\delta b,ij}$
into the off-diagonal cross-power $P^{\rm clust}_{\varepsilon,ij}(k)$ exactly as it does the diagonal.

\subsection{Sub-Poisson stochasticity (halo exclusion)} \label{sec:subP}
For a real, mass-selected sample this assembly-bias excess is not the whole story: the same discreteness that gives the Poisson floor also enforces \emph{halo exclusion} (two haloes cannot overlap), which \emph{lowers} the small-scale shot noise. Following Baldauf et al.~\cite{baldauf2013}
we add their exclusion correction,
\begin{equation}
\label{eq:Pexcl}
P_\varepsilon^{\rm excl}(k)=-V_{\rm excl}\,W_R(k)-b_1^2\,V_{\rm excl}\,[W_R\!*\!P_{mm}](k),
\qquad V_{\rm excl}=\tfrac43\pi R_{\rm excl}^3,
\end{equation}
with $R_{\rm excl}$ the exclusion radius and $W_R$ a top-hat. Baldauf et al. measure $R_{\rm excl}$ directly from the halo--halo correlation function and find it at $0.8$ times the $\xi_{hh}$ maximum. The fitted values are provided in Table I therein. 

For the comparison in Section~\ref{sec:pkcomp} we adopt a slightly different choice. We multiply their found exclusion radii by a factor $1.25$ corresponding to the exact peak of the correlation function. As shown in Table~\ref{tab:rexcl}, this choice is equivalent to the Lagrangian radius $R_\mathrm{excl}\approx R_L$ for the lowest mass bins. Furthermore, it matches the stochasticity at high-mass bins better, as we will explain later.

We generalise halo exclusion to the off-diagonal exactly as in Baldauf et al.: $b_1^2\to b_{1,i}b_{1,j}$ and $R\to R_{ij}=(R_i+R_j)/2$ in the same smoothed window \eqref{eq:Pexcl}, while the pure Poisson term $\delta_{ij}/\bar n$ sits on the diagonal alone. Assembled, the full stochasticity matrix~\eqref{eq:stochsum} now carries both non-Poisson pieces: its diagonal is the $\sigma_{ii}(k)$ that Fig.~\ref{fig:pkcross} tests against Baldauf et al.'s Fig.~7, and its off-diagonals are the cross terms we turn to in the comparison below.

\subsection{Comparison with previous studies}
\label{sec:pkcomp}

\begin{table}
\centering
\begin{tabular}{lccccc}
\hline
bin & $M_h\,[10^{13}h^{-1}\Msun]$ & $R_{\rm excl}^{\rm B13}$ & $1.25\,R_{\rm excl}^{\rm B13}$ (this work) & $R_L$ & ratio \\
\hline
I    & 1.14 & 2.6 & 3.25  & 3.40 & 0.96 \\
II   & 1.27 & 2.8 & 3.50  & 3.52 & 0.99 \\
III  & 1.42 & 2.8 & 3.50  & 3.66 & 0.96 \\
IV   & 1.62 & 3.0 & 3.75  & 3.82 & 0.98 \\
V    & 1.89 & 3.6 & 4.50  & 4.02 & 1.12 \\
VI   & 2.25 & 3.9 & 4.88  & 4.26 & 1.14 \\
VII  & 2.80 & 4.2 & 5.25  & 4.58 & 1.15 \\
VIII & 3.72 & 4.9 & 6.12  & 5.04 & 1.21 \\
IX   & 5.65 & 5.8 & 7.25  & 5.79 & 1.25 \\
X    & 16.6 & 8.7 & 10.88 & 8.30 & 1.31 \\
\hline
\end{tabular}
\caption{Exclusion radii (in $h^{-1}$Mpc) bin by bin for Baldauf et al.'s \cite{baldauf2013} tabulated Lagrangian values $R_{\rm excl}^{\rm B13}$ (their convention: $0.8\times$ the maximum of $\xi_{hh}$), the $\xi_{hh}$-maximum convention adopted here (multiplying $R_{\rm excl}^{\rm B13}$ by $1.25$), and the true Lagrangian radius $R_L=(3M_h/4\pi\bar\rho)^{1/3}$; the last column is $1.25\,R_{\rm excl}/R_L$. At the light bins the $\xi_{hh}$ maximum coincides with the Lagrangian radius to a few per cent; toward the massive end the exclusion zone grows to $1.3\,R_L$.}
\label{tab:rexcl}
\end{table}

\begin{figure}
\centering
\includegraphics[width=\columnwidth]{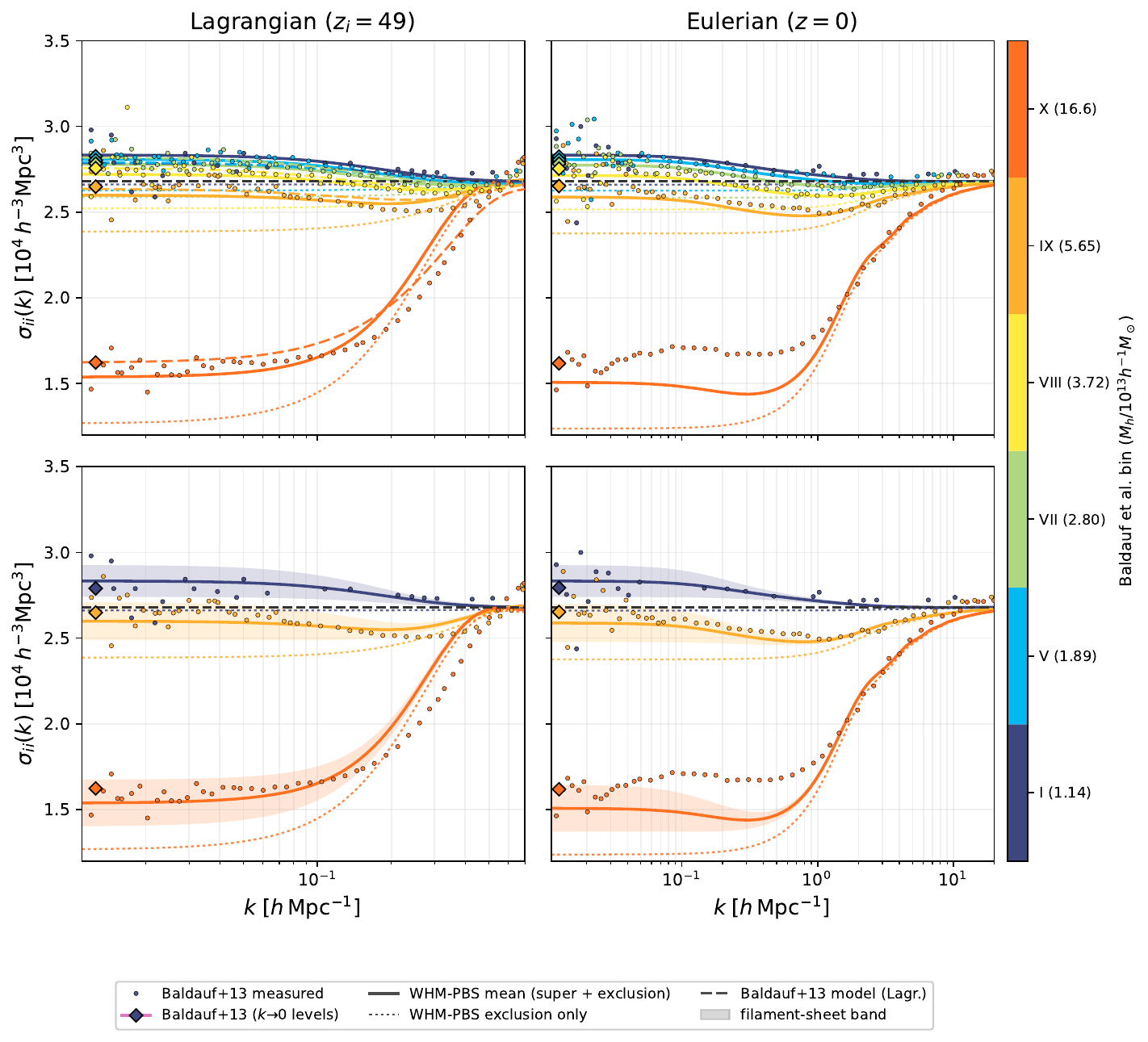}
\caption{The \whmpbs bias-scatter stochasticity with halo exclusion, against \emph{both} frames of Fig.~7 of Baldauf et al.~\cite{baldauf2013} --- the traced-back Lagrangian field ($z_i=49$, left) and the evolved Eulerian field ($z=0$, right) --- in their per-bin colours (WMAP3, their Table-1 bins and biases). \emph{Top row:} the model headline prediction, the \emph{mean} of the filament- and sheet-host predictions (solid), against Baldauf et al.'s own Table-1 model (dashed, Lagrangian only; they give no evolved-field model) and their measured points, for bins I, V and VII--X. \emph{Bottom row:} bins I, IX and X alone --- the mean, the parameter-free filament--sheet \emph{band} ($1\sigma=$ half their difference), and the pure halo exclusion (dotted; $1/\bar n$ plus exclusion, super-Poisson removed); every panel carries the measured points of the bins it shows. The heavy horizontal dashed line is the Poisson floor $1/\bar n=26800\,h^{-3}\mathrm{Mpc}^3$. The host scatter uses the spherical top-hat window (Lagrangian); in the Eulerian frame the host is the WHM nonlinear matter power~\cite{whm} convolved with the collapsed filament and sheet windows.
}
\label{fig:pkcross}
\end{figure}

Figure~\ref{fig:pkcross} shows an apples-to-apples comparison between the Baldauf et al.~\cite{baldauf2013} model and data in Fig.~7 therein and our model. It shows the Lagrangian field ($z_i\!=\!49$, left) and the evolved Eulerian field ($z=0$, right), with their measured stochasticity matrix diagonal elements $\sigma_{ii}(k)$ for a selection of mass bins. The data is extracted using the WebPlotDigitizer tool on their published figure. We use their conventions throughout: their WMAP3 cosmology, their ten halo-mass bins and Eulerian biases (their Table~1),  and their \emph{smoothed} exclusion window [their Eq.~(43), the erf-in-$\log r$ transition that gives the clean return to the Poisson floor at high $k$], with the exclusion radii at the $\xi_{hh}$-maximum convention of Section~\ref{sec:subP}. 

Here, we have the freedom to choose as the halo's host either the filament or the sheet. By closure, both readings share the same mean bias. What differs is the scatter of hosts, which is broader at the sheet level. Since the super-Poisson amplitude is set by the scatter rather than by the mean, each choice yields a single, fully specified stochasticity curve, and the two curves differ: the filament host sets the lower limit, the sheet host the upper. Both track the data reasonably well, so we treat the pair as bracketing the model freedom: the region between the two curves is our parameter-free theory prior band (lower row of Fig.~\ref{fig:pkcross}, with $1\sigma =$ half their difference), and their mean is the central prediction shown in both rows.

The top-left (Lagrangian) panel shows the joint $P_\varepsilon$ prediction including the clustering and halo exclusion parts developed in sections \ref{sec:superP} and \ref{sec:subP}. The top-right (Eulerian) panel evolves \emph{both} ingredients into the collapsed frame: the host scatter is convolved with the \emph{non-linear} web--halo model (\whm~\cite{whm}) matter power spectrum. Also, we replace the host-tophat-windows $W_{R_\star}$ by their collapsed counterparts (Equations (36)-(38) of \cite{whm}). 
Further, the exclusion window becomes the collapsed halo's NFW window $u_{\rm NFW}(k\,|\,R_{\rm vir},c)$ {with the Bullock et al. \cite{bullock2001} concentration, at a higher radius ($R_L/3$, as found by Baldauf et al.) than the Eulerian virial radius. 

The diagonal stochasticity $\sigma_{ii}(k)=1/\bar n+P_\varepsilon^{\rm net}$ reproduces the structure of their measurement in both frames: a modest super-Poisson excess at low mass,  a sub-Poisson suppression in the most massive bin, where exclusion dominates, and a return to the Poisson floor $1/\bar n$ at high $k$, which in the Eulerian frame is delayed and broadened by the collapsed halo's profile. The amplitude of the excess is set by the host scatter $\mathrm{Var}(b_1)$. 

The dotted curves give the pure halo-exclusion
prediction (Poisson floor plus exclusion, with the super-Poisson host term switched off); the gap up to the full solid curve is the super-Poisson host scatter, which the light bins require to reach their measured level and which the exclusion alone cannot supply. The Lagrangian panel also carries, as dashed curves, Baldauf et al.'s \emph{own} model --- their smoothed-step fit with a quadratic bias tuned to each bin, computed here from their Table~1. The \whmpbs prediction tracks the data about as closely across the bins without fitting an effective nonlinear bias $b_2$ as in their case. At the largest mass bin they find an imaginary value for $b_2$, because of an underestimation of the halo exclusion effect that requires the (Super-Poisson) clustering component to be negative. This is the reason why we adapted larger halo exlcusion radii throughout. The impact of this is reduced at lighter mass bins, where the clustering effect dominates. The most massive bin's level alone prefers a slightly smaller $1.2\,R_{\rm excl}$, which we do not adopt here but flag as a target for a more complete exclusion treatment.

\begin{figure}
\centering
\includegraphics[width=\columnwidth]{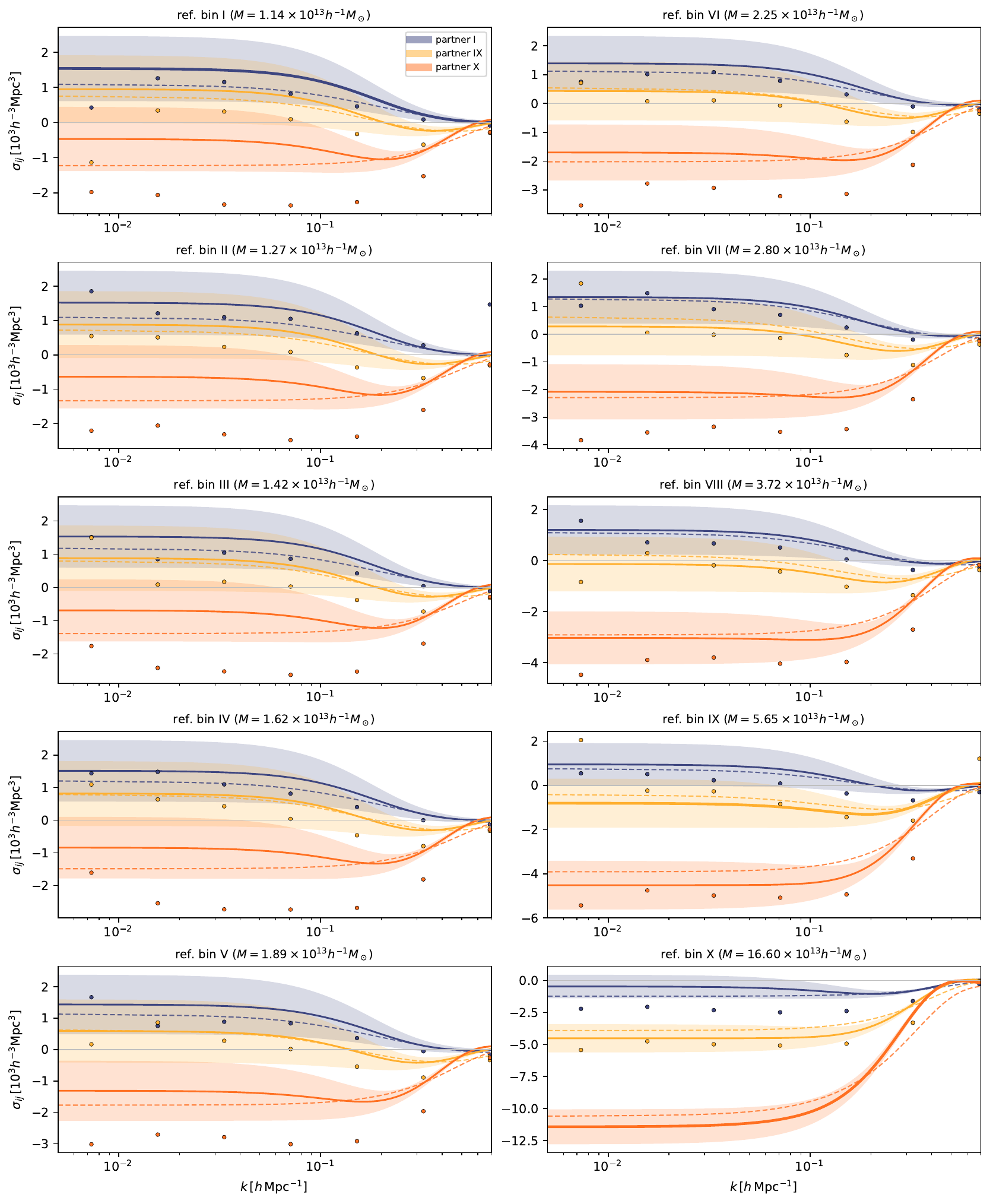}
\caption{The full \whmpbs stochasticity matrix, all ten reference bins (one panel per bin, mass increasing), shown as the filament--sheet \emph{band} in Lagrangian space and the $1\sigma$ correposnding to half their difference) with the mean (solid) for three partner bins --- I, IX and X --- against Baldauf et al.'s measured entries (points) and their Table-1 fit (dashed). Colour keys the partner bin. The band brackets the measured cross terms across the matrix up to bin IX.}
\label{fig:stochmatrixall}
\end{figure}

We next consider the off-diagonals of the stochasticity matrix. A cross term between two distinct bins carries no self-pairs. Hence there is no Poisson floor, so it is the cleaner probe of the host scatter's origin, that the \whmpbs predicts (Section~\ref{sec:superP}).

Figure~\ref{fig:stochmatrixall} shows the resulting cross terms in the Lagrangian conventions of Fig.~\ref{fig:pkcross}. Against a low-mass reference bin every entry (except the highest mass bin) is super-Poisson, strongest for its neighbours (the implied bias-fluctuation correlation is $r_{ij}\simeq0.98$) and decaying slowly toward distant partners ($r_{1,10}\simeq0.6$). Against the most massive bin the cross exclusion outweighs the shared-host clustering and the whole X row, I$\times$X through IX$\times$X, turns \emph{negative}. 

To compress our results in an efficient way, we 
consider the eigenvalues of the stochasticity matrix, plotted as a function of $k$ in Fig.~\ref{fig:stochmatrix}. The ten-bin matrix has ten eigenvalues, and if the bins were independent Poisson tracers all ten would sit at $1/\bar n$. Whatever structure the noise has beyond discreteness induces eigenvalues departing from the Poisson floor. Diagonalising $\sigma_{ij}(k\!\to\!0)$ exposes exactly two (Fig.~\ref{fig:stochmatrix}) departures: one enhanced eigenvalue, one suppressed, and eight at the Poisson floor. Again, the \whmpbs filament and sheet predictions set the lower and upper envelope, and we take their means as our central prediction. We overplot the simulation data measured in Baldauf et al. \cite{baldauf2013} (digitised from Fig. 12 therein). Overall, they are in very good agreement with the \whmpbs band.

Finally, we discuss how our model predictions convert to prior bands to be used on EFT analyses of the halo power spectrum. Since the clustering observations take place at late time, we need to operate in the Eulerian frame (right column of figure \ref{fig:pkcross}). In that frame, the diagonal stochasticity elements are nearly flat up to the scales probed with EFT ($k_\mathrm{max}\approx 0.2 h/\mathrm{Mpc}$), such that we only consider a scale-independent shift of the shotnoise amplitude. 
We assume a Gaussian centered at the mean between the filament and sheet prediction and adopt their difference as the $\pm \sigma$-range. This prior will be used in Section~\ref{sec:demo}.

\begin{figure}
\centering
\includegraphics[width=0.75\columnwidth]{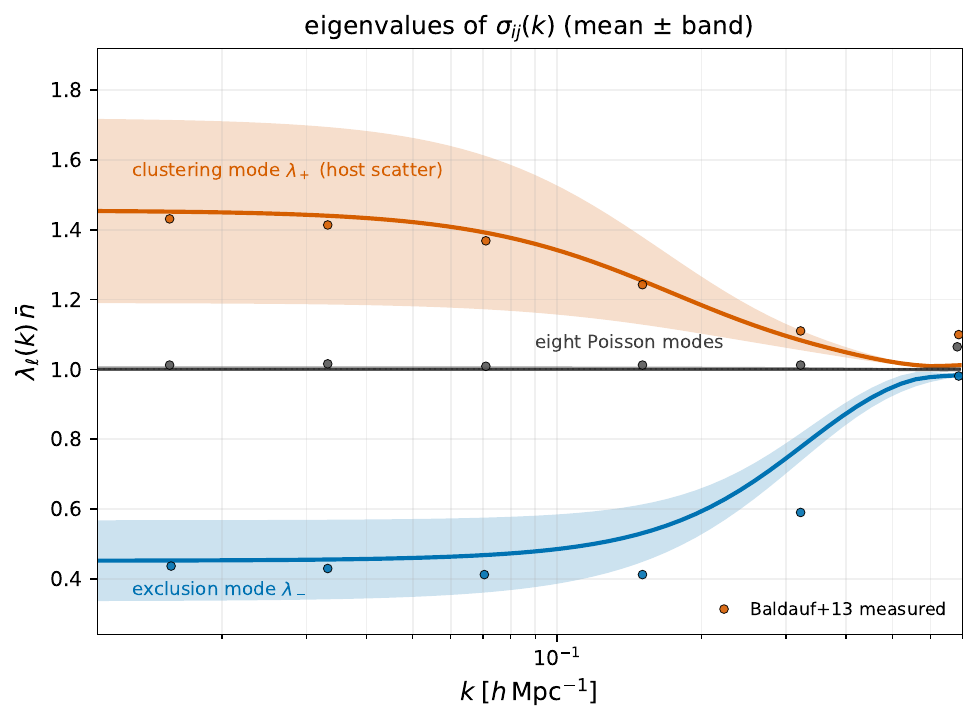}
\caption{The two non-Poisson eigenvalues $\lambda_\pm(k)/(1/\bar n)$ of the \whmpbs stochasticity matrix $\sigma_{ij}(k)$ (spherical top-hat, Lagrangian), as the \emph{mean} of the filament and sheet hosts with the filament--sheet band ($1\sigma=$ half their difference): the clustering mode $\lambda_+$ (host scatter) above and the exclusion mode $\lambda_-$ below the eight Poisson modes at unity.}
\label{fig:stochmatrix}
\end{figure}

%% file: sections/assembly_bias.tex
\section{\texorpdfstring{Assembly bias: the bias--concentration inversion and the tidal prior}{Assembly bias: the bias-concentration inversion and the tidal prior}}
\label{sec:assembly}
The width of the inherited-bias distribution has now been interpreted in two of its three ways: as a selection \emph{uncertainty} it became the prior band of Section~\ref{sec:predictions}, and as a \emph{random} field it became the halo stochasticity of Section~\ref{sec:pk}. We now turn to the third: the \emph{correlation} between a halo's environment and its internal structure.
 
At fixed mass, halo clustering depends systematically on secondary properties -- weakly at high mass, but by up to a factor of ${\sim}2$ in bias at low mass -- an effect known as \emph{assembly bias} \citep{ShethTormen2004,gao2005,gao2007}. One example is the bias--concentration correlation (BCC): the sign of $\partial b_1/\partial c|_{\Mh}$ is \emph{positive} at low mass (more concentrated haloes are more strongly clustered) and \emph{negative} at high mass \citep{wechsler2006,gao2007,jing2007}, inverting near the mass at which the mass fraction in haloes has a maximum, $m_\ast\!\approx\!2\times10^{13}\,h^{-1}\Msun$ \citep{paranjape2018}. 

To be clear from the start: within the \whmpbs framework developed here we can only make statements about how a halo's bias correlates with its \emph{external} environment; by itself the framework cannot predict correlations with \emph{internal} halo properties. Establishing the connection between environment and concentration, spin or mass-accretion history requires an additional model of how a halo's internal structure responds to its environment, which we leave to future work. We stress that any residual assembly bias that does \emph{not} correlate with environment at all lies outside the reach of the \whmpbs formalism.

Here we therefore focus on the environment side alone and ask to what extent the \whmpbs formalism may be useful for assembly-bias \emph{model-building}. 
Concretely, we show that the mass at which the BCC changes sign and the environment side of its amplitude are explained by host-population mixing, without any model linking concentration to environment, and we deduce the coupling strength such a concentration--environment model would have to supply, to match the measured correlation coefficient (Section~\ref{sec:assembly_amplitude}). In the tidal sector the corresponding assembly-bias signal is directly measured in simulations, and we use it to calibrate the selection term of the tidal prior of Section~\ref{sec:bs2pred} (Section~\ref{sec:tidal_assembly}).

\subsection{\texorpdfstring{The bias--concentration correlation from population mixing}{The bias-concentration correlation from population mixing}}
\label{sec:assembly_mixing}
\label{sec:assembly_amplitude}
 Haloes of mass $\Mh$ do not have \emph{one} environment:  they occupy a \emph{distribution} of filament hosts $P(\Mf|\Mh)$ [Eq.~\eqref{eq:conditional}, Fig.~\ref{fig:cond}c], and by the environment-averaged bias \eqref{eq:envbias} it inherits the bias of that host. The mixing picture assumes that a halo's secondary properties track the host's tidal character: a halo whose host is much more massive than itself ($\Mf/\Mh$ large) sits in an anisotropic, filament-dominated, tidally arrested environment, while one whose host is barely more massive sits in a near-isotropic, node-like environment. These are the two populations Paranjape, Hahn and Sheth~\citep[PHS18,][]{paranjape2018} separate by the measured anisotropy of the local tidal field: the anisotropic population carries a \emph{positive} BCC -- tidally arrested haloes that retain dense cores while residing in strongly clustered regions, the non-accreting subpopulation identified by \cite{dalal2008} and attributed to tidal truncation of accretion and to splashback/ejected orbits \citep{hahn2009,borzyszkowski2017,mansfield2020} --, while the isotropic population carries the \emph{negative} peaks-curvature signal \citep{dalal2008}. In \whmpbs the host mass ratio $\Mf/\Mh$ is the excursion-set measure of this tidal character.

The all-halo BCC is the population-weighted average of the two, and it inverts where they balance, that is, where the \emph{typical} (median) halo's host crosses the turnaround threshold
\begin{equation}
  \frac{\Mf}{\Mh}=\frac{\bar\rho_{\rm ta}}{200\bar\rho}\Big(\frac{R_{\rm ta}}{R_{200b}}\Big)^{3} \simeq \frac{5.5}{200}4^3 = 1.76~,
  \label{eq:turnaround}
\end{equation}
where we assumed the turnaround scale $R_{\rm ta}\simeq4\,R_{200b}$ and overdensity $\bar\rho_{\rm ta}=5.5\bar\rho$. This is precisely the scale on which
PHS18 (Appendix~B therein) smooth the tidal field to characterise each halo's environment, and at which the environment--bias correlation peaks.\footnote{In particular, they find that their tidal anisotropy parameter $\alpha_R$ shows the strongest correlation with bias when smoothed at $R=R_\mathrm{ta}$.}

\begin{figure}
\centering
\includegraphics[width=\columnwidth]{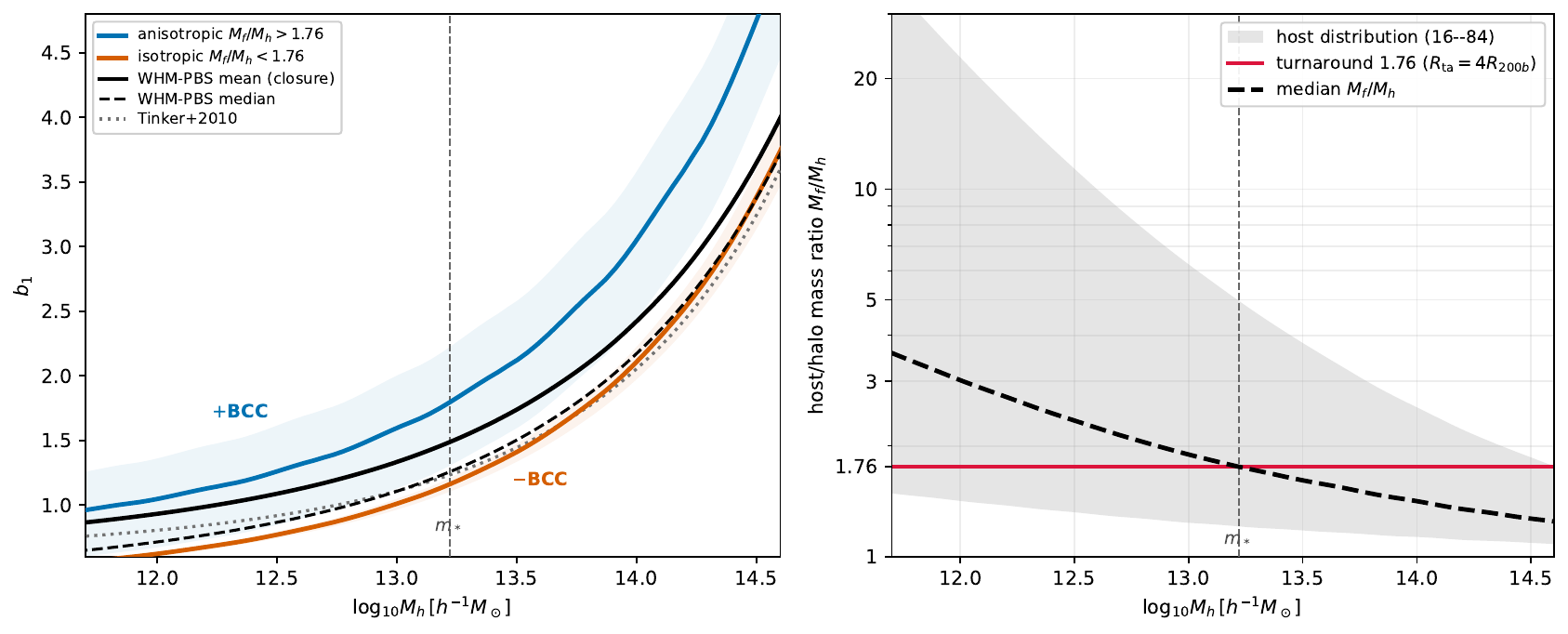}
\caption{The bias--concentration inversion as population mixing. \emph{Left:} the host-ratio population split.%
The isotropic population is tight and tracks the Tinker et al.~\cite{tinker2010} fit (grey dotted), while the broad anisotropic tail pulls the \whmpbs\ mean (closure, black solid) above the median (black dashed). The two bands reproduce the node and filament sequences of PHS18~\cite{paranjape2018} (their Fig.~14, left), namely the tight isotropic (node-like) and broad anisotropic (filament) classes. This is the excursion-set analogue of their tidal-anisotropy dissection, in which the large-scale bias rises with the anisotropy at fixed mass.%
}%
\label{fig:bcc_mixing}
\end{figure}

The physical intuition is as follows: haloes whose environment is restricted to a region \emph{within} $R_{\rm ta}$ are actively attracting all matter surrounding it. So they are not really part of a larger-scale filament, but instead an (isotropic) ``proto-halo'' growing further in mass with time.
Evaluating the exact conditional $P(\Mf|\Mh)$ for the filament host, the median host ratio falls through this turnaround mass ratio threshold $1.76$ at
\begin{equation}
  b_1^{\ast}\simeq1.5\qquad(\Mh = m_{\ast}\simeq1.7\times10^{13}\,h^{-1}\Msun),
  \label{eq:inversion}
\end{equation}
within ${\sim}5\%$ in $b_1$ of the PHS18 inversion
($m_\ast\!=\!2\times10^{13}$, corresponding to $b_1^\ast\!\approx\!1.55$ in our case).
Figure~\ref{fig:bcc_mixing} (right) shows the crossing directly: below $m_\ast$ the median host lies above the turnaround line ($\Mf/\Mh>1.76$, anisotropic, positive BCC), while above $m_\ast$ it falls below it (isotropic, negative BCC).

The same host distribution also predicts not only the \emph{crossing}, but also the environment side of the BCC \emph{amplitude}. The bias contrast between the two populations, namely the inherited host bias of the anisotropic half minus that of the isotropic half, $\Delta b_1=\langle b_1\rangle_{\Mf/\Mh>1.76}-\langle b_1\rangle_{\Mf/\Mh<1.76}$, is parameter-free (Fig.~\ref{fig:bcc_mixing}, left, the excursion-set analogue of their Fig.~14).
At the inversion, where the split is at the median, it  approximates to the \emph{scatter} of the bias distribution at fixed mass,
\begin{equation}
  \Delta b_1\;\simeq\;\sigma(b_1|\Mh)\;\simeq\;0.6\qquad(b_1\simeq1.5).
  \label{eq:bccamp}
\end{equation}
Resolved into the two populations, the scatter itself reproduces the morphological split of
PHS18~\cite{paranjape2018} (their Fig.~14, left): the isotropic half is a tight,
low-bias sequence tracking the standard mass--bias relation, while the anisotropic half is a
broad, high-bias tail. These are the node and filament classes, recovered here from the host
distribution alone and shown as the two scatter bands of Fig.~\ref{fig:bcc_mixing}.
This $\mathrm{Var}(b_1|\Mh)$ is the \emph{same} quantity behind the beyond-Poisson
stochasticity of Section~\ref{sec:pk}: the assembly-bias amplitude and the stochasticity are the two observable faces of one predicted variance.

Equation~\eqref{eq:bccamp} fixes the \emph{bias} side of the correlation. The observable strength depends in addition on how concentration responds to the environment. Writing the concentration at fixed mass as $c=\bar c(\Mh)+\lambda\,\delta b_1+\varepsilon$, where $\delta b_1 = \sigma(b_1)$ is the environmental bias fluctuation, $\lambda=\partial\ln c/\partial b_1$ the tidal coupling, and $\varepsilon$ the intrinsic (formation-time) scatter, the bias--concentration correlation coefficient is
\begin{equation}
  r(b_1,c)=\frac{\lambda\,\sigma(b_1)}{\sqrt{\lambda^2\sigma^2(b_1)+\sigma_\varepsilon^2}}.
  \label{eq:rbcc}
\end{equation}

So far we have only motivated the \emph{sign} of the tidal coupling $\lambda$, but we were not able to predict its \emph{amplitude}. The latter -- and a model for the intrinsic scatter $\sigma_\varepsilon$ -- would be necessary to predict the LHS of Eq.~\eqref{eq:rbcc}.
Here, we take a different route and motivate the value $\lambda$ should obtain to match simulation-based measurements. First, the intrinsic concentration scatter has been measured in \cite{bullock2001} to about $\sigma_\varepsilon\!=\!\sigma(\ln c)\!\approx\!0.3$ ($0.18\,$dex). Then, to match the small correlation $|r_S(b_1,c)|{\lesssim}0.1$ measured by PHS18~\cite{paranjape2018} (their Fig.~15) through \eqref{eq:rbcc} \emph{requires} a weak coupling $\lambda\!\approx\!0.1$.\footnote{Note that this coupling is different from the quantity $\lambda_{S}$ used in Eqs.~\eqref{eq:tidsel} and~\eqref{eq:tid-priorband} (Section~\ref{sec:tidal_assembly}), which quantifies the assembly bias in the \emph{tidal} bias, while here we are only referring to the linear bias.}

This reasoning provides a first step towards developing a full analytic model for assembly bias, which we leave to future work.

\subsection{Assembly bias in the tidal sector}
\label{sec:tidal_assembly}

The tidal prior of Section~\ref{sec:bs2pred} carries a second ingredient the density prior does not --- a term for how far a host-correlated selection can move a sample off the broken $\bso(b_1)$ relation. That term is assembly bias seen in the tidal sector, and we bound it here. Weighting the environment average \eqref{eq:tid-envbias} by a selection $w(x)$ on any internal property $x$ (luminosity, colour, concentration, spin) at fixed mass gives the effective tidal bias
\begin{equation}
\label{eq:tidsel}
\bso^{S}=\frac{\big\langle w\,\bso^{L}(\Mh|\Mf)\big\rangle_{\Mf|\Mh}}{\langle w\rangle_{\Mf|\Mh}}
\equiv\big\langle\bso^{L}\big\rangle_{\Mf|\Mh}+\lambda_{S}\,\sigma_{\rm band},
\end{equation}
which defines $\lambda_{S}$, the fraction of the per-object band $\sigma_{\rm band}$ a selection reaches. A host-blind selection has $\lambda_{S}=0$ and sits on the closure mean. A selection departs from it only through the correlation between what it selects on and the pair's tidal weight.

Lazeyras et al.~\cite{lazeyrasbarreira2021} split each mass bin into quartiles of concentration or spin, which displaces the Eulerian tidal bias off the mass-binned relation by $\simeq0.26$--$0.44\,\sigma_{\rm band}$ across $b_1\simeq2$-$4.8$, on average $\lambda_{S}\simeq0.35$. Basically, this is the most a selection can do, and we adopt it as the conservative tidal-prior width [Eq.~\eqref{eq:tid-priorband}]. Consistently, the same splits leave the \emph{density} relation $b_2(b_1)$ in place (Section~\ref{sec:bN}).

%% file: sections/demo.tex
\section{Demonstration: WHM-PBS priors in a real-space EFT inference}
\label{sec:demo}

The prior set developed across the manuscript --- the selection band [Eq.~\eqref{eq:priorband}], the tidal band [Eq.~\eqref{eq:tid-priorband}], the derived cubic bias [Eq.~\eqref{eq:bgamma3}], and the stochasticity prediction of Section~\ref{sec:pk} --- is meant to be used. Because these relations are nearly cosmology-universal (Section~\ref{sec:predictions}), the bias priors are evaluated as fixed functions of the sampled $b_1$ at every step of the chain rather than recomputed per cosmology. This section demonstrates it end-to-end in a controlled setting: a one-loop EFT fit of the real-space halo power spectrum for a synthetic, DESI-like suite of three tracer bins. We restrict this demonstration to real space, because there the one-loop nuisance sector mostly corresponds to the \whmpbs predictions presented in this paper: the bias set $\{b_1,b_2,\bso,\bGa\}$ (Section~\ref{sec:predictions}), the stochastic amplitude $\mathrm{SN}_0$ (Section~\ref{sec:pk}), and one counterterm $\alpha\,k^2P_{11}$ (the only ingredient not predicted in this work). The velocity sector that redshift space adds (Finger-of-God counterterms, anisotropic shot noise) is physics the \whmpbs does not yet model. The redshift-space application is hence left for future work. Also, since we have not modelled the galaxy-halo connection yet, we will assume throughout that galaxies are perfect tracers of haloes, so we use the terminology `galaxies' and `haloes' interchangeably.

\paragraph{Setup.} The synthetic mock dataset is generated at a $\Lambda$CDM truth
$(\omega_c,h,\ln 10^{10}A_s)=(0.11933,0.6766,3.0448)$ using the velocileptors one-loop,
IR-resummed EPT power spectrum~\cite{chen2020} for three bins mimicking DESI LRG and ELG samples (redshift $z=0.51,\,0.71,\,1.32$; linear bias $b_1=1.93,\,2.08,\,1.45$; density $\bar n=5,4,5\times10^{-4}\,(h/\mathrm{Mpc})^{3}$; and volume $V=3.7,\,5.5,\,11\,(\mathrm{Gpc}/h)^{3}$), with truth nuisances drawn from the \emph{external} literature, not from the \whmpbs: $b_2$ from the Lazeyras et al.\ fits~\cite{lazeyras2016}, $\bso$ from the Abidi--Baldauf fits~\cite{abidi2018}, $\bGa$ determined by Eq.~\eqref{eq:bgamma3}. The
stochasticity is a single number per bin: the $k\to0$ shot-noise \emph{level} measured by
Baldauf et al.~\cite{baldauf2013}, interpolated at each bin's host mass ($\mathrm{SN}_0-1/\bar n=+184,\,-1009,\,+1380\,(\mathrm{Mpc}/h)^3$; the same $k\to0$ diamonds in the right panel of Fig.~\ref{fig:pkcross}). 

We generate the DESI-inspired halo-halo auto ($P_{hh}(k)$) and halo-matter cross ($P_{hm}(k)$) power spectrum data sets, in the wavenumber range $0.02 < k [h/\mathrm{Mpc}] < 0.2$ with binsize $\Delta k = 0.005 h/\mathrm{Mpc}$. 

Because mock and model coincide, $\chi^2(\mathrm{truth})=0$ exactly in every configuration whose parameter space contains the truth: whatever a fit recovers away from the truth is pure inference geometry --- prior volume, marginalisation, projection --- which is this section's subject. The covariance is diagonal Gaussian, $\mathrm{Cov}=2\,(P+1/\bar n)^2/N_k$, with $k\in[0.02,0.20]\hMpc$. The linear parameters $\{\alpha,\mathrm{SN}_0\}$ (and, where free, $\bGa$) are marginalised analytically. Posteriors are obtained using nested sampling implemented in Nautilus~\cite{lange2023}.

\paragraph{Configurations.} We investigate six different prior choices, each of which corresponds to one sampling run. In all cases, the parameters $(\omega_c,h,\ln 10^{10}A_s,b_1,\alpha)$ are varied freely assuming flat priors. We adopt the \emph{`Free'} setup (i) with flat, wide priors on $b_2$, $\bso$, $\bGa$ and $\mathrm{SN}_0$. The  \emph{\whmpbs} setup (ii) adopts the $b_2(b_1)$ band of Eq.~\eqref{eq:priorband}, the tidal band of Eq.~\eqref{eq:tid-priorband}, the Eq.~\eqref{eq:bgamma3} relation, and the Gaussian $\mathrm{SN}_0$ prior from section \ref{sec:pkcomp}. The remaining four are different variants of keeping the same nuisance parameters \emph{fixed}: either at the \emph{truth} (iii), at the \whmpbs prediction \emph{centres} (iv), or at the truth displaced by $\pm1\sigma$ of the \whmpbs widths, (v) and (vi).

\begin{figure*}[t]
\centering
\includegraphics[width=\textwidth]{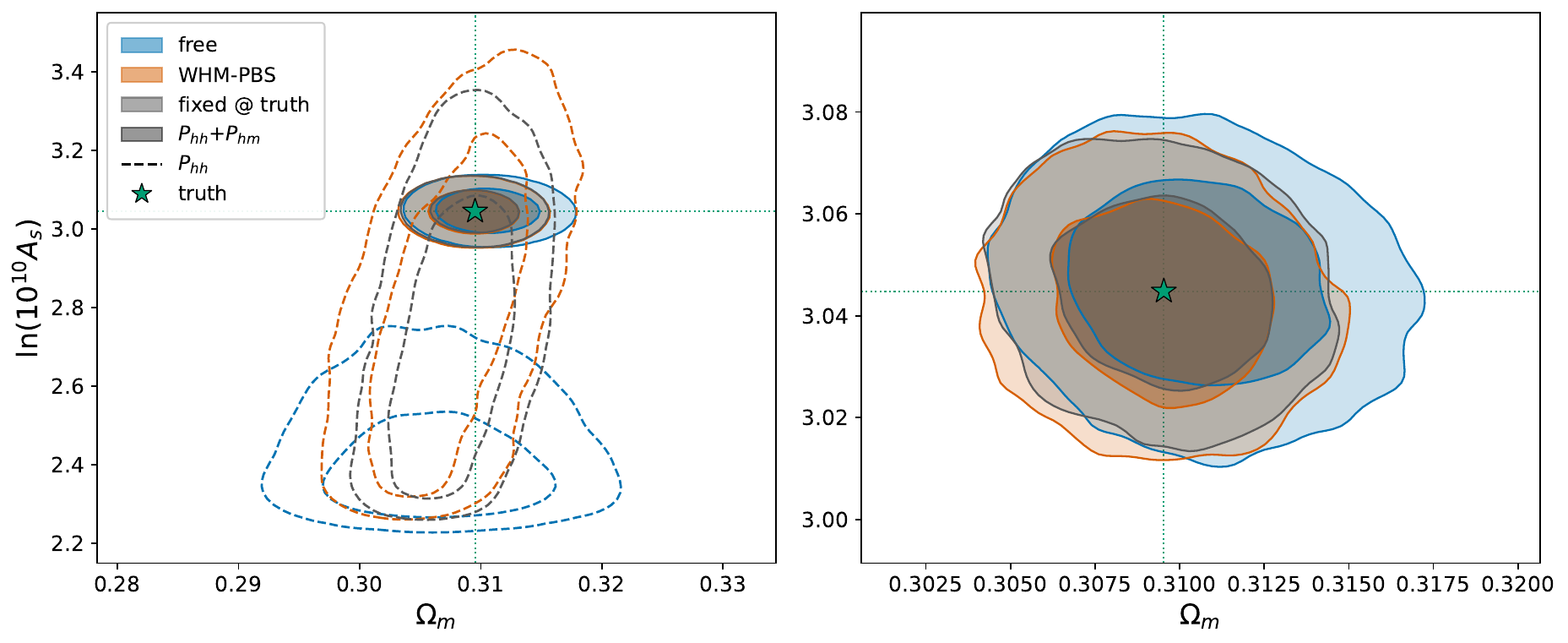}
\caption{We show the $\ln(10^{10}A_s)$--$\Omega_m$ degeneracy for the \emph{free} fit (blue), the full \whmpbs\ prior set (orange) and the truth-fixed nuisances (grey), with the truth displayed with a green star. Dashed, 
unfilled contours display auto power spectrum fits only, filled solid contours the joint auto+cross power spectrum fits. The left panel shows both cases, whereas the right panel zooms into the range of the posteriors from the joint fits.}
\label{fig:rsdemo-degen}
\end{figure*}

\paragraph{Results.} Fitting the auto-spectrum $P_{hh}(k)$ alone, the `\emph{free}' setup results in a spurious low-$\ln 10^{10}A_s$, low-$\Omega_m$ constraint, driven by the strong $b_1^2A_s$ degeneracy that is broken only weakly by one-loop corrections (${\sim}7\%$ of $P$ at $k=0.1\hMpc$). As shown in the left panel of Fig.~\ref{fig:rsdemo-degen} (dashed lines), either adopting the \whmpbs priors or fixing them ameliorate this prior volume effect partially, such that the truth resides within the $1-\sigma$ constraint. The fundamental problem of the amplitude degeneracy within the halo real-space power spectrum is well known \citep{pezzotta2021,euclidRealSpace2024} and it has been shown that it can be remedied by including the cross spectrum $P_{hm}(k)$. Indeed, since the latter scales linearly in the bias, the $b_1^2A_s$ degeneracy is broken, such that the data becomes constraining enough and all posteriors recover the truth, although the `\emph{free}' case still shows a residual bias of $\sim 0.4\sigma$ in $\Omega_m$.

A full quantitative comparison of all cases (i)-(vi) is shown in Table~\ref{tab:rsdemo}. It includes the observed shifts in cosmological parameters in units of the standard derivation and the Figure of Bias (FoB). We find that the \whmpbs priors are constraining enough to mitigate projection effects and flexible enough to yield unbiased (FoB$<1$) constraints. Fixing the nuisance parameters to values $1\sigma$ away from the truth, however, yields up to $\sim 3\sigma$ biases in cosmological parameters.

Finally, Fig.~\ref{fig:rsdemo-degen} shows the full triangle plot of the \emph{`free'} and \emph{\whmpbs} cases, where the upper half is a zoom into the more constraining auto+cross posteriors. Overplotted is the \whmpbs prior in grey in the relevant panels, showing how it helps to break the amplitude degeneracy.

\begin{table}[t]
\centering
\caption{Here we summarise the results of the real-space demonstration at $k_{\max}=0.2\hMpc$: the figure of bias $\mathrm{FoB}=\sqrt{\Delta^T C^{-1}\Delta/3.53}$ over $(\Omega_m,h,\ln 10^{10}A_s)$, with the posterior-mean shifts of all three in units of $\sigma$, for the auto spectrum $P_{hh}$ (top) and the joint $P_{hh}+P_{hm}$ (bottom). \emph{Fixed} rows are delta-function pins of $(b_2,\bso,\bGa,\mathrm{SN}_0)$ with $b_1$ sampled: at the truth, at the \whmpbs prediction
centres, and at the truth displaced by $\pm1\sigma$ of the \whmpbs widths.}
\label{tab:rsdemo}
\begin{tabular}{lcccc}
\hline
treatment & FoB & $\Delta\Omega_m/\sigma$ & $\Delta h/\sigma$ & $\Delta\ln A_s/\sigma$ \\
\hline
\multicolumn{5}{l}{\emph{auto spectrum $P_{hh}$}}\\[1pt]
free (wide) & 3.15 & $-0.57$ & $-0.92$ & $-5.83^{\dagger}$ \\
\textbf{\whmpbs} & \textbf{0.49} & $\mathbf{-0.51}$ & $\mathbf{-0.08}$ & $\mathbf{-0.90^{\dagger}}$ \\
fixed @ truth & 0.71 & $-0.52$ & $-0.07$ & $-1.33^{\dagger}$ \\
fixed @ \whmpbs\ centres & 0.20 & $-0.01$ & $+0.18$ & $-0.36$ \\
fixed @ truth $+1\sigma_{\rm WHM}$ & 1.97 & $-1.10$ & $-1.28$ & $-2.50^{\dagger}$ \\
fixed @ truth $-1\sigma_{\rm WHM}$ & 1.53 & $+1.29$ & $+1.55$ & $+0.34$ \\
\hline
\multicolumn{5}{l}{\emph{with the halo--matter cross $P_{hh}+P_{hm}$}}\\[1pt]
free (wide) & 0.41 & $+0.42$ & $+0.27$ & $+0.07$ \\
\textbf{\whmpbs} & \textbf{0.15} & $\mathbf{-0.05}$ & $\mathbf{+0.01}$ & $\mathbf{-0.13}$ \\
fixed @ truth & 0.01 & $-0.01$ & $+0.00$ & $-0.00$ \\
fixed @ \whmpbs\ centres & 1.23 & $-1.53$ & $-0.34$ & $+0.47$ \\
fixed @ truth $+1\sigma_{\rm WHM}$ & 11.05 & $+0.34$ & $+1.62$ & $+2.95$ \\
fixed @ truth $-1\sigma_{\rm WHM}$ & 9.24 & $+0.02$ & $-1.40$ & $-2.46$ \\
\hline
\multicolumn{5}{l}{\footnotesize $^{\dagger}$ prior-dominated: the $P_{hh}$ posterior piles at the $\ln 10^{10}A_s$ prior edge, so $\sigma$ is bound-limited.}\\
\end{tabular}
\end{table}

\begin{figure*}[tp]
\centering
\includegraphics[width=\textwidth]{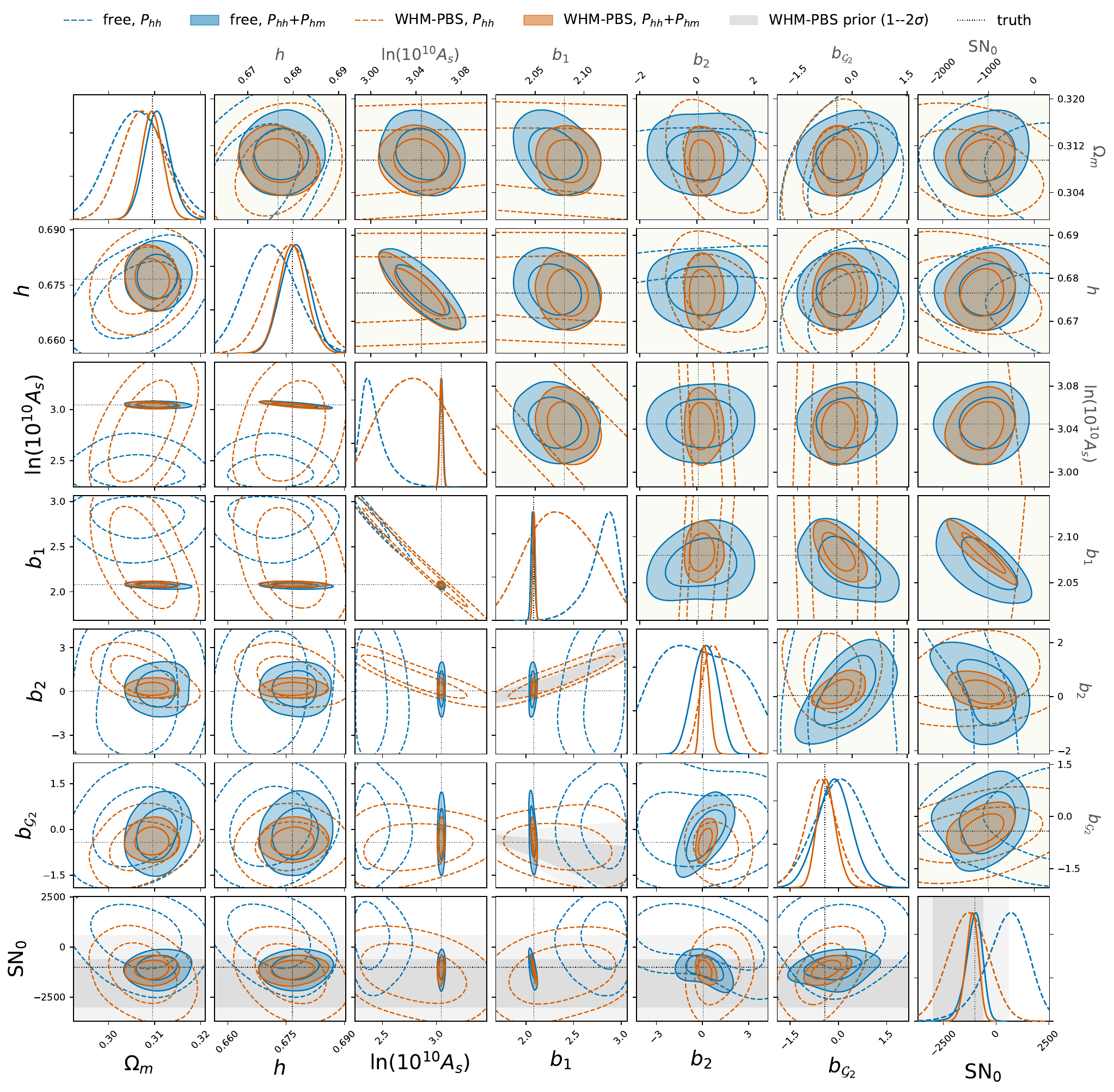}
\caption{Here, we show the real-space demonstration for the LRG2 bin ($z=0.71$; $68/95\%$ contours), over the cosmology $(\Omega_m,h,\ln 10^{10}A_s)$ and that bin's nuisances $(b_1,b_2,\bso,\mathrm{SN}_0)$, for the \emph{free} broad-prior fit (blue) and the full \whmpbs prior set (orange). Dashed (unfilled): the auto spectrum $P_{hh}$; filled: the joint $P_{hh}+P_{hm}$ with the halo--matter cross. \emph{Lower triangle:} full ranges. \emph{Upper triangle:} the same pairs zoomed onto the $P_{hh}+P_{hm}$ region (tinted; the top and right axes carry the zoom scale). Grey bands mark the informative \whmpbs priors: the $\mathrm{SN}_0$ mean/band level ($1$--$2\sigma$, wherever $\mathrm{SN}_0$ appears) and, shown only against $b_1$, the $b_2(b_1)$ and $\bso(b_1)$ bands. For $P_{hh}$ only the amplitude $\ln 10^{10}A_s$ slides down the $b_1$--$A_s$ valley to the prior edge for \emph{both} fits (the $(\ln 10^{10}A_s,b_1)$ panel). The halo--matter cross spectrum breaks the degeneracy and collapses every parameter onto the truth (dotted). The \whmpbs bias bands pin $b_2$ and $\bso$ where the free fit is unconstrained, while $\mathrm{SN}_0$ is data-dominated in both fits --- sitting at the truth well inside its wide prior.}
\label{fig:rsdemo}
\end{figure*}

%% file: sections/conclusions.tex
\section{Conclusions}
\label{sec:conclusions}

We have developed the Web--Halo Model Peak--Background Split (\whmpbs), an analytic theory in which the large-scale bias of a dark-matter halo is inherited from the cosmic-web environment it occupies: the bias of a halo is the bias of its host filament, inside its host sheet, averaged over the conditional mass function --- the environment-averaged bias of Eq.~\eqref{eq:envbias}. The change of viewpoint can be put in one sentence. Because the host mass is random at fixed halo mass, halo bias is not a number but a \emph{distribution}, and the nuisance parameters of galaxy clustering --- the higher-order and tidal biases, the shot noise, the assembly-bias amplitude --- are not independent free functions but \emph{moments of one host distribution} $P(\Mf|\Mh)$, computed from the cosmic web with a single calibrated barrier for each morphology. Its mean is the familiar bias; its variance is the source of nearly everything else. Our main results are:

\begin{enumerate}

\item \textbf{Halo bias is inherited, and the relation closes.} The bias of a halo equals the environment-averaged bias of its host \eqref{eq:envbias}. Computed as the response to an additive background shift it closes exactly and reproduces the \emph{shape} of the Tinker et al.~\cite{tinker2010} calibration; the mean is level-independent (the nested sheet$\to$filament$\to$halo hierarchy and the two-level halo-in-filament form agree to ${\lesssim}0.6$\,pp). The tidal bias follows from the same barriers' response to a long-wavelength shear \eqref{eq:tid-uncond} and obeys the same hierarchical closure.

\item \textbf{Bias is a skewed distribution, and its variance is the deliverable.} \whmpbs predicts the full $P(b|\Mh)$: the mean reproduces the bias relations, the median lies below it, and the $N$-body measurements are bracketed by the two. That variance enters the observables in three ways which are \emph{not} independent. As a \emph{selection uncertainty} it becomes a prior band on the bias relations (result~3); as a \emph{random field} it sources a super-Poisson stochasticity in the halo power spectrum which, together with halo exclusion, reproduces the super- to sub-Poisson trend of Baldauf et al.~\cite{baldauf2013}; and as a \emph{correlation} with halo structure it becomes assembly bias (result~4). The latter two are the same variance seen random and correlated; and the correlation one feeds back into the first, because it is the tidal-sector assembly-bias signal that calibrates the tidal prior.

\item \textbf{Density and tidal bias relations, and the priors they carry.} The same exact response predicts $b_2(b_1)$ and $b_3(b_1)$ on the universal relations of Lazeyras et al.~\cite{lazeyras2016}, and $\bso(b_1)$ spanning the $N$-body measurements \citep{lazeyras2018,abidi2018,zennaro2022,modi2017}. Because the density biases are slaved to a single host property they carry no scatter at fixed $b_1$: a sample slides \emph{along} them, which is why they are universal and survive the assembly-bias splits of Lazeyras et al.~\cite{lazeyrasbarreira2021}. An aggregation rule \eqref{eq:aggregation} fixes what a finite sample adds on top, lifting it off the per-object curve giving the tight selection prior band \eqref{eq:priorband}. The tidal relation is genuinely \emph{broken} --- a background shear couples to the halo's own mass, not the host alone --- so it scatters across $b_1$, and its typical-host track (the fixed-mass median at its own median $b_1$) runs along the measured relation (RMS $0.09$, against $0.51$ for the mean).

\item \textbf{An assembly-bias inversion that also calibrates the tidal prior.} The bias--concentration correlation is not the zero of a bias coefficient but population mixing: the all-halo signal inverts where the typical host crosses the turnaround threshold $\Mf/\Mh\simeq1.76$ ($R_{\rm ta}=4R_{200b}$, the scale on which Paranjape et al.~\cite{paranjape2018} measure the tidal anisotropy), placing the inversion at $b_1\simeq1.5$ ($\Mh^\ast\simeq1.7\times10^{13}\simeq m_\ast$) to within ${\sim}5\%$, with no parameter tuned to assembly bias. Its environment side is the predicted scatter $\sigma(b_1)$, the same $\mathrm{Var}(b_1)$ as the stochasticity. The concentration- and spin-split measurements of Lazeyras et al.~\cite{lazeyrasbarreira2021} that quantify this correlation in the \emph{tidal} sector also fix the tidal prior's selection term: an outright ranking on an internal property displaces the tidal bias by at most $0.35\,\sigma_{\rm band}$, which we adopt as the conservative width (Section~\ref{sec:tidal_assembly}). This is a different, and several times larger, coupling than the density-sector $r_S\simeq0.1$ of Paranjape et al., which is why the density prior carries no such term. Assembly bias is thus not a separate corollary but the calibration of one half of the prior set.

\item \textbf{The priors demonstrated end-to-end.} In a synthetic real-space EFT inference of three DESI-like tracer bins (Section~\ref{sec:demo}), broad nuisance priors alone drift the shape parameters down the $b_1$--$A_s$ amplitude valley, which the \whmpbs density-bias band cures on its own, recovering $\Omega_m$ and $h$ at the projection floor. Pinning the nuisances at their predicted centres looks safe in the auto spectrum, but the halo--matter cross exposes the pinned centre, and a $1\sigma$ pin calibration error fails outright, whereas the same error handed to the \emph{band} is absorbed by its width. The value demonstrated is not sharper central values but \emph{calibrated widths}.

\end{enumerate}

\textbf{The \whmpbs prior set (a practitioner's summary).} For a full-shape analysis the deliverables condense to four rules. \emph{(1) Density biases:} take $b_2(b_1)$ Gaussian, centred on the mass-binned mean relation (or any simulation-based calibration) with width $\sigma_{b_2}=\mathrm{Var}(b_1|\Mh)\simeq0.4$ (a $b_2$-axis width the quadratic relation sets from the $b_1$ variance), truncated below at the per-object host envelope exactly $1\sigma$ under the centre \eqref{eq:priorband}. For $b_3$ centre on the mean (or external calibration) with a width of order its lift and no hard floor. \emph{(2) Tidal bias:} take $\bso(b_1)$ Gaussian, centred on the mean relation with $\sigma^{2}=[\bso^{\rm mean}-\bso^{\rm host}]^{2}+[0.35\,\sigma_{\rm band}]^{2}$ \eqref{eq:tid-priorband} --- an envelope between the mean and the host relation, plus the tidal-sector selection term set to the property-ordered ceiling $0.35\,\sigma_{\rm band}$ calibrated in Section~\ref{sec:tidal_assembly} and adopted as the conservative choice. \emph{(3) Cubic bias:} slave $\bGa$ to $(b_1,\bso^{L})$ via \eqref{eq:bgamma3}. \emph{(4) Stochasticity:} put a Gaussian prior on the shot-noise level $\mathrm{SN}_0$, centred on the mean of the filament and sheet $k\to0$ levels (super-Poisson at low mass, sub-Poisson at high) with half their spread as the $1\sigma$ width, anchored at $z=0$. Build the noise of a multi-tracer analysis from the matrix of Section~\ref{sec:pkcomp}.

%\red{\textbf{On cosmology-universality and its limits.} The prior relations above are functions of $b_1$, not of cosmology: within the Markovian walk every \whmpbs quantity depends on the power spectrum only through $S=\sigma^2(M)$, which the mapping to $b_1$ eliminates, so the $b_N(b_1)$ relations are universal across cosmology (Section~\ref{sec:predictions}), not merely across redshift, and may be computed once and applied at the sampled $b_1$. In $\Lambda$CDM the only residual is the collapse threshold $\dc(\Omega_m(z))$ in the barrier, at the sub-percent level. Beyond $\Lambda$CDM the universality can weaken through four handles: (i)~$\dc$ acquiring stronger or scale-dependent running (dynamical dark energy; massive neutrinos and modified gravity, where the collapse threshold turns scale-dependent); (ii)~a deformed power-spectrum shape (neutrino free-streaming, warm dark matter, a running $n_s$) activating the non-Markovian spectral-slope dependence \citep{mussosheth2012,scs2013} that the Markovian walk sets to zero; (iii)~scale-dependent growth breaking the $\sigma^2(M,z)=D^2(z)\,\sigma^2(M,0)$ separability behind the redshift relabelling, which also forces the cold-plus-baryon variance $\sigma_{cb}(M)$ in place of the total-matter one \citep{castorina2014}; and (iv)~the ellipsoidal-collapse barrier parameters, calibrated to $\Lambda$CDM $N$-body simulations, drifting in exotic cosmologies. All four are small against the current prior widths but bound the regime in which the universality holds, and frame the survey-grade companion.}

The \whmpbs formalism presented here bears several limitations. The analytic mean sits ${\sim}18\%$ above the measured linear bias \citep{lazeyras2016,tinker2010}. A barrier refit to the bias as well as the mass function is the natural route, left to future work. The offset does not propagate into the falsifiable predictions, which live in the $b_N(b_1)$ relations where $b_1$ is marginalised, not in the absolute $b_1(M)$. The tidal response \eqref{eq:tid-uncond} is the linear peak--background response to an effective shear barrier; as Desjacques et al.~\cite{desjacques2018tidal} note, such barriers do not satisfy the full microscopic bias consistency relation, and our response omits the correlated-step and curvature contributions of the complete non-local bias \citep{scs2013}.

We envision several directions to extend the \whmpbs formalism in future work. Currently, it predicts the large-scale bias of \emph{haloes in real space}. The extension to observed galaxies in redshift space requires a halo-occupation distribution and a velocity model. This would allow for a survey-grade application similar to Section~\ref{sec:demo} but on a realistic prior set. Also, the \whmpbs captures only environmental dependence as assembly bias and says nothing about halo properties (formation time, spin, shape) that do not reduce to host mass and web level. And turning the predicted environment--bias correlation into a prediction for a specific secondary observable, as we do for concentration in Section~\ref{sec:assembly}, requires an external model linking that observable to the environment. Other natural next steps would be the extension to the correlated-step (non-Markovian) walk and generalisation to primordial non-Gaussianity.

%% file: sections/appendices.tex
\section{Exact first crossing versus the truncated series}
\label{app:taylor}

\begin{figure*}[t]
\centering
\includegraphics[width=\textwidth]{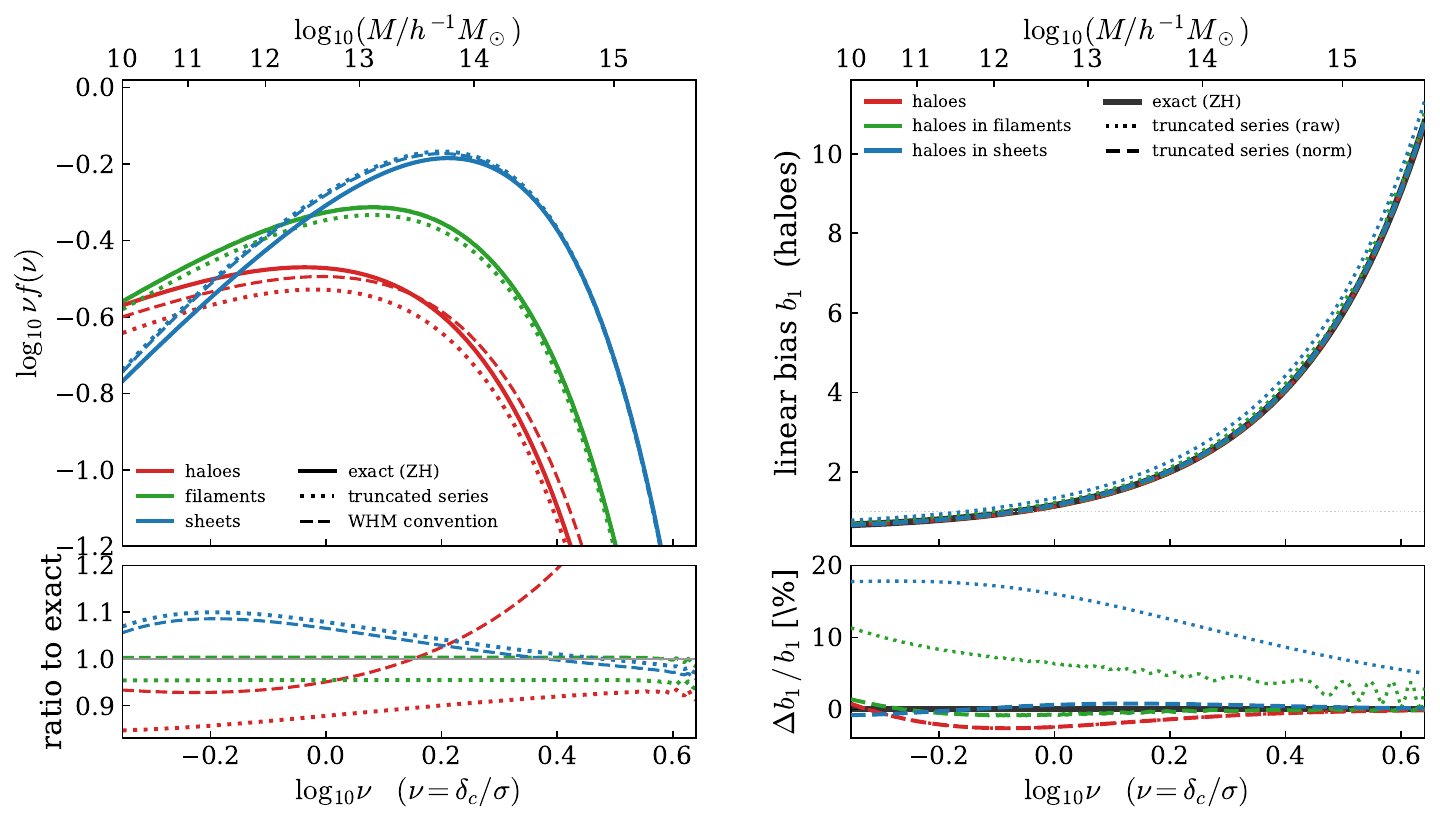}
\caption{Exact ZH first crossing versus the truncated Sheth and Tormen~\cite{shethtormen2002} Taylor series, as functions of peak height $\nu=\delta_c/\sigma$ (lower axis) and halo mass (upper axis), down to $10^{10}\,h^{-1}\Msun$. \emph{Left, top:} the unconditional first-crossing multiplicity $\nu f(\nu)$ (with $\nu=\delta_c/\sigma$, the convention of Fig.~2 of the WHM paper) for haloes (red), filaments (green) and sheets (blue), all with $a=0.707$. There are three curves per morphology: the exact solver (solid), the truncated Shen series with its original amplitude $A=0.639$ (dotted), and the WHM convention (dashed), see main text for details.
\emph{Left, bottom:} the corresponding ratios to the exact solver. \emph{Right:} the linear bias $b_1$ of a halo for three host treatments -- the halo alone (red), in a filament (green), in a sheet (blue) -- with the truncated host average shown two ways: \emph{self-normalised} (dashed, divided by the host-weight sum $\int f_{\rm par}F\,\mathrm dS_p$) and \emph{raw} (dotted, divided instead by the unconditional halo crossing $f_h(S_h)$ that closure says the convolution should equal). Their gap is the Chapman--Kolmogorov violation of the truncated series. The exact solver is level-independent, the \emph{single} black curve in the background. The self-normalised curves hug it to ${\lesssim}1\%$ -- the renormalisation hides the violation -- whereas the raw curves depart by its full size: nothing for the halo alone (no host), ${\sim}7\%$ for the filament, and up to ${\sim}17\%$ for the
sheet (lower-right, residual).}
\label{fig:zhtaylor}
\end{figure*}

In the early stages of this work the first-crossing distribution $f(S)$ was evaluated with the truncated Taylor series of Sheth and Tormen~\cite{shethtormen2002}, a low-order expansion of the moving-barrier crossing rate. The series is convenient, but it is not normalised: its integral $\int f\,\mathrm dS$ does not return the true crossing probability $P_{\rm cross}$, and it mis-normalises each morphology by a \emph{different} amount. Evaluated on a common range of $S$, it over-counts the sheet barrier by ${\sim}6\%$, reproduces the filament normalisation to better than a percent, and under-counts the steeply rising halo barrier by ${\sim}8\%$ (Table~\ref{tab:zhtaylor}), so the three abundances are no longer mutually consistent. The exact ZH solver of Appendix~\ref{app:zh} carries no such error, returning $P_{\rm cross}$ by construction. Figure~\ref{fig:zhtaylor} compares the two solvers directly, down to $10^{10}\,h^{-1}\Msun$: the left panel shows the three unconditional mass functions, the right the linear bias they predict.

Adopting the exact solution is crucial, because the whole construction rests on the Bayesian inversion $P(\Mf|\Mh)=N(\Mh|\Mf)\,n_f(\Mf)/n_h(\Mh)$ [Eq.~\eqref{eq:conditional}], which integrates to unity only if the abundances composing it are mutually consistent [the Chapman--Kolmogorov composition \eqref{eq:CK}]. With the un-normalised series this inversion does not close, and the early pipeline had to force-normalise each conditional by hand to recover a probability. The non-normalisation of these conditional mass functions was initially the showstopper of the project, a problem that the exact solver removes.

The consequence for the bias is shown in the right-hand panel of Fig.~\ref{fig:zhtaylor}, for three treatments of the host: the halo alone, the halo in a filament, and the halo in a sheet. By closure the exact solver gives a \emph{single}, level-independent curve (black): the host-averaged $\langle b_1\rangle_{\Mf|\Mh}$ equals the directly computed halo bias to ${\lesssim}0.1\%$ whether the halo sits in a filament, in a sheet, or nowhere (the two- and three-level means agree to $\{3.5\times10^{-4},\,2.7\times10^{-3},\,2.0\times10^{-2}\}$ for $N=1,2,3$; Section~\ref{sec:envbias}). The panel also shows \emph{how far} the truncated series breaks closure. A host average can be formed two ways: \emph{self-normalised} (dashed), dividing by the host-weight integral $\int f_{\rm par}(S_p)\,F(S_h|S_p)\,\mathrm dS_p$, or \emph{raw} (dotted), dividing instead by the unconditional halo crossing $f_h(S_h)$. The two agree if that convolution equals $f_h(S_h)$, which is precisely the Chapman--Kolmogorov closure, so their ratio measures the violation. For the exact solver they coincide; for the truncated series the self-normalised curves still hug the exact to ${\lesssim}1\%$ -- the renormalisation hides the violation, which is why the early pipeline missed it -- while the raw curves expose it in full: nothing for the halo alone, ${\sim}7\%$ for the filament, and up to ${\sim}17\%$ for the sheet, growing as the host barrier drops further below the halo barrier. The self-normalisation also re-introduces a spurious level dependence; the exact, closed theory needs neither it nor any rescaling.

The left-hand panel of Fig.~\ref{fig:zhtaylor} compares, for each morphology, the exact first crossing (solid) with the truncated Shen series at its original amplitude $A=0.639$ (dotted) and with the WHM convention (dashed): ST99 for the halo, and the Shen series at the mass-conserving eq.~(28) amplitudes therein for filaments and sheets. The truncation itself is mild: it reproduces the filament, over-counts the (downward) sheet barrier by up to ${\sim}10\%$ at low mass, and under-counts the steeply rising halo barrier. The eq.~(28) adjustment then shifts the filament up by $5.2\%$ and the sheet down by $1.3\%$ relative to the original Shen amplitude -- the small mass-conservation rescaling, which is the only difference between the dotted and dashed curves for those two morphologies. The sheet shift carries limited consequence, since the sheet enters the WHM only through its sub-dominant collapse-order term, with little leverage on the matter power spectrum. For the halo the WHM uses not the Shen-barrier series -- which diverges unphysically as $\nu\to0$ -- but the ST99 mass function of Sheth and Tormen~\cite{shethtormen1999}, which tracks the exact solver to a few percent, departing only at the
high-mass tail where the steeper Shen barrier suppresses the most massive haloes.

For these reasons we use the exact ZH first crossing throughout. It is non-negative, exact, and normalised to $P_{\rm cross}$ by construction, so the conditional mass functions integrate to unity, the closure of Section~\ref{sec:envbias} holds with no imposed renormalisation, and the mean bias is genuinely level-independent.

\begin{table}[t]
\centering
\caption{The unconditional crossing probability $\int f\,\mathrm dS$ of the truncated Sheth and Tormen~\cite{shethtormen2002} Taylor series against the exact ZH solver, on the same barrier parameters, cosmology, and range of $S$. The truncated series mis-normalises each morphology by a different amount, whereas the exact solver returns $P_{\rm cross}$ by construction.}
\label{tab:zhtaylor}
\begin{tabular}{lcc}
\hline
 & ZH & truncated series \\
\hline
sheets    & $0.99$ & $1.05$ \\
filaments & $0.93$ & $0.94$ \\
haloes    & $0.75$ & $0.70$ \\
\hline
\end{tabular}
\end{table}

\section{Solving the Zhang--Hui equation}
\label{app:zh}
All first crossings are computed from the Volterra equation \eqref{eq:fcross} with kernels \eqref{eq:kernels}. Discretising the variance on a uniform grid $S_i=i\,\Delta S$ ($i=0,\dots,n$) and applying the trapezoidal rule to the integral turns \eqref{eq:fcross} into a \emph{lower-triangular} linear system, in which $g_2(S_i,S_j)$ couples $f_i$ only to $f_{j\le i}$, solved by forward substitution,
\begin{equation}
\label{eq:zh-forward}
f_i=\frac{g_1(S_i)+\sum_{j=1}^{i-1}\big(\Delta_{i,j}+\Delta_{i,j+1}\big)\,f_j}{1-\Delta_{ii}},
\end{equation}
with $\Delta_{ij}\equiv\tfrac12\Delta S\,g_2(S_i,S_j)$, starting from $f_1=g_1(S_1)/(1-\Delta_{11})$. The diagonal needs care: as $S'\!\to\!S$ the propagator $P_0(B(S)-B(S'),S-S')$ carries an integrable $1/\sqrt{S-S'}$ singularity, so we replace $\Delta_{ii}$ by the interval average
\begin{equation}
\label{eq:zh-diag}
\begin{aligned}
\Delta_{ii}&=\frac12\int_{S_i-\Delta S}^{S_i}\!\!g_2(S_i,S')\,\mathrm dS'\\
&=\frac12\int_0^{\sqrt{\Delta S}}\!\!g_2\big(S_i,\,S_i-t^2\big)\,2t\,\mathrm dt,
\end{aligned}
\end{equation}
where the substitution $S'=S_i-t^2$ cancels the singularity and the remaining smooth integrand is evaluated by $32$-point Gauss--Legendre quadrature. The solution is exact up to the grid, strictly non-negative, and normalised to $\int f\,\mathrm dS=P_{\rm cross}$ with no imposed re-normalisation. This is the property that makes the conditional \eqref{eq:conditional} integrate to unity and the environment-averaged bias \eqref{eq:envbias} close, and that the truncated Sheth and Tormen~\cite{shethtormen2002} series does not share. The conditional crossing $F(S_h|S_f)$ uses the same solver on the shifted barrier $\tilde B(s)=B_h(S_f+s)-B_f(S_f)$.

Finally, the peak--background-split bias \eqref{eq:pbsbias} is the response of this solution to the rigidly shifted barrier $B(S)\to B(S)-\dl$, taken by five-point finite differences in $\dl$ (to $N$-th order for $b_N$); the tidal bias \eqref{eq:tid-uncond} is the analogous response to the slope-weighted shift of Appendix~\ref{app:tidal}.

\section{The exact ($\delta$-response) bias}
\label{app:pbsbias}

In the sharp-$k$ excursion set $\delta(S)$ is Brownian in $S=\sigma^2(M)$ and a patch collapses at the first up-crossing of the barrier $B(S)$, with first-crossing density $f(S)$ and $\nu=\dc/\sqrt S$. A long-wavelength mode $\dl$ raises the density uniformly at every scale, $\delta(S)\mapsto\delta(S)+\dl$; since a patch then crosses when $\delta(S)+\dl=B(S)$, the background is equivalent to lowering the barrier uniformly to $B(S')-\dl$. The Lagrangian bias is the normalised first-crossing response \eqref{eq:pbsbias-f}, restated here as the starting point of the kernel expansion,
\begin{equation}
\label{eq:appdef}
b_N^{L}(S)=\frac{1}{f(S)}\,\frac{\partial^N\hat f(S;\dl)}{\partial\dl^N}\bigg|_{\dl=0},
\qquad
\hat f(S;\dl)\equiv f[\,B-\dl\,](S),
\end{equation}
where the bracket notation $f[\,\cdot\,](S)$ stresses that the crossing density is a \emph{functional} of the whole barrier function: it depends on $B(S')$ at \emph{all} scales $S'$, through the kernels of \eqref{eq:fcross}. We take $N=1$; higher orders are identical.

Because the barrier enters $f$ at every scale, the response is carried by the functional (Fr\'echet) derivative of $f$, which is \emph{defined} as the coefficient of the term linear in an arbitrary barrier perturbation $\eta(S')$:
\begin{equation}
\label{eq:appvar}
f[\,B+\epsilon\eta\,](S)=f[B](S)+\epsilon\int_0^S\!\mathrm{d}S'\,K(S,S')\,\eta(S')+O(\epsilon^2),
\qquad
K(S,S')\equiv\frac{\delta f(S)}{\delta B(S')}.
\end{equation}
This is the origin of the kernel \emph{and} of the integral: a barrier perturbation is a \emph{function} of the scale $S'$, so to linear order its effect on $f(S)$ is the kernel $K(S,S')$ integrated against the perturbation over all scales. The walk is causal, so $K(S,S')=0$ for $S'>S$ and the integral runs over $[0,S]$.

The background is the special case of a \emph{uniform} perturbation, $\eta(S')=-1$ at every scale with $\epsilon=\dl$, because $B-\dl$ lowers the barrier by $\dl$ everywhere. Putting $\eta=-1,\ \epsilon=\dl$ into \eqref{eq:appvar} expands $\hat f$ explicitly in powers of $\dl$,
\begin{equation}
\label{eq:appexp}
\hat f(S;\dl)=f[\,B-\dl\,](S)=f[B](S)\;-\;\dl\int_0^S\!\mathrm{d}S'\,K(S,S')\;+\;O(\dl^2),
\end{equation}
a constant (the unperturbed crossing) plus a term linear in $\dl$ whose coefficient is the integrated kernel. Differentiating in $\dl$ and setting $\dl=0$ removes the constant and the $O(\dl^2)$ remainder and selects this linear coefficient,
\begin{equation}
\label{eq:appderiv}
\frac{\partial\hat f(S;\dl)}{\partial\dl}\bigg|_{\dl=0}=-\int_0^S\!\mathrm{d}S'\,K(S,S'),
\end{equation}
with $K$ evaluated at the unperturbed barrier $B$. Dividing by $f(S)$ gives the direct ($\delta$-response) bias,
\begin{equation}
\label{eq:appdirect}
b_1^{L}(S)=-\frac{1}{f(S)}\int_0^S\!\mathrm{d}S'\,K(S,S').
\end{equation}
Each ingredient now has a definite origin: the integral over $S'$ is the linear term of the functional expansion \eqref{eq:appexp}, summing how $f$ responds to lowering the barrier at \emph{each} scale; the minus sign is the rate $-1$ at which the background lowers it; and the ``$|_{\dl=0}$'' is what selects this linear response, with the kernel taken at the unperturbed barrier. Equation~\eqref{eq:appdirect} reduces to the Press--Schechter $b_1^L=(\nu^2-1)/\dc$ for the constant barrier $B=\dc$. This additive response is the consistent definition of the bias for a \emph{moving} barrier; it coincides with the log-derivative $-\partial\ln(\nu f)/\partial\dc$ of a universal mass function only for the constant spherical-collapse barrier ($\partial B/\partial\dc=1$), as treated in detail by Desjacques et al.~\cite{desjacques2018} and verified against simulations by Lazeyras et al.~\cite{lazeyras2016}. We validate \eqref{eq:appdirect} against Press--Schechter (constant barrier: $b_1,b_2,b_3$ to $<0.1\%$) and against the environment-averaged closure \eqref{eq:envbias}: the marginalisation $\langle b_f\rangle_{\Mf|\Mh}=b_h$ holds to $+0.03\%$, the consistency the host-averaging of Section~\ref{sec:envbias} requires. The direct response is what closes the theory and carries the tidal sector and the assembly-bias inversion (Section~\ref{sec:assembly}).

\section{The exact tidal response and its closure}
\label{app:tidal}
This appendix records the kernel form of the tidal response defined in Section~\ref{sec:tidal},
its three-level telescoping, and the closure that follows. Throughout, $K(S,S')=\delta f(S)/\delta
B(S')$ is the first-crossing response kernel introduced for the density sector in
Appendix~\ref{app:pbsbias} [Eq.~\eqref{eq:appvar}].

\paragraph*{The kernel (convolution) form.} According to \eqref{eq:tid-shift} a long-wavelength shear shifts the barrier by the slope-weighted, \emph{non-uniform} profile $\delta B=\tfrac32\,s_b^2\,\mathrm dB/\mathrm dS$. This is the same functional perturbation as in the density sector but with a scale-dependent profile $\propto\mathrm dB/\mathrm dS$, so inserting it into the expansion \eqref{eq:appvar} and differentiating at $s_b^2=0$ turns the response definition \eqref{eq:tid-uncond} into a convolution of the kernel with the barrier slope,
\begin{equation}
\label{eq:tid-kernel}
\bso^{L}(S)=\frac{3}{2}\,\frac{1}{f(S)}\int_0^S\!\mathrm dS'\,K(S,S')\,
\frac{\mathrm dB}{\mathrm dS}(S').
\end{equation}
This is the tidal counterpart of the direct density bias \eqref{eq:appdirect}: there the shift is the uniform $\eta=-1$, the constant leaves the integral, and it collapses to $-\bar K(S)\equiv-f^{-1}\!\int_0^S K\,\mathrm dS'=b_1^L(S)$; here the profile $\mathrm dB/\mathrm dS$ is scale-dependent and stays inside the integral. The $3/2=\mathrm dS/\mathrm d\langle s^2\rangle$ is the Jacobian of the shear-variance relation $\langle s^2\rangle=\tfrac23 S$ \eqref{eq:tid-shift}, not a free factor. For the ellipsoidal barrier \eqref{eq:barrier} the slope is analytic, $\mathrm dB/\mathrm dS=\alpha\,[B(S)-\sqrt a\,\dc]/S$, namely $\alpha$ times the moving (shear) part of the barrier over $S$, so the tidal weight is literally the shear part of the barrier, and a \emph{constant} (spherical) barrier ($\beta=0$, $\mathrm dB/\mathrm dS=0$) gives $\bso^{L}=0$.

\paragraph*{Why the local product fails.} Replacing the slope by its endpoint value and pulling it out of the integral gives the local (single-scale) approximation
\begin{equation}
\label{eq:tid-local}
\bso^{L,\,\rm local}(S)=\frac{3}{2}\,\bar K(S)\,\frac{\mathrm dB}{\mathrm dS}(S)
=-\frac{3}{2}\,b_1^L(S)\,\frac{\mathrm dB}{\mathrm dS}(S),
\end{equation}
exact only for a constant slope. For the moving halo barrier the slope varies across the support of $K$, so the local form under-counts $\bso$: the exact ratio (exact/local) tends to unity only at large $\nu^2$ and reaches $\sim3$--$5$ at low $\nu^2$, with a sign change near $\nu^2\!\sim\!1$, precisely where the local product also violates the closure below. Equation~\eqref{eq:tid-kernel}
must therefore be used in full.

\paragraph*{Three-level telescoping and closure.} Unlike the uniform density shift, the shear does \emph{not} leave the conditional crossings invariant: it shifts every barrier by its own slope, so each barrier difference $B_h-B_f$ and $B_f-B_s$ changes and the conditional crossings $N(\Mh|\Mf)$ and $N(\Mf|\Ms)$ respond. Differentiating the nested Chapman--Kolmogorov composition $\nh=\int\!\!\int N(\Mh|\Mf)\,N(\Mf|\Ms)\,n_s(\Ms)\,\mathrm d\Mf\,\mathrm d\Ms$ under the consistent shift, the Lagrangian tidal bias of a halo in a given filament-in-sheet telescopes into one term per level,
\begin{equation}
\label{eq:tid-nested}
\begin{aligned}
\bso^{L}(\Mh|\Mf,\Ms)=\tfrac32\Big[&\,
\mathcal K_s\big[\tfrac{\mathrm dB_s}{\mathrm dS}\big]
+\mathcal K_{f|s}\big[\tfrac{\mathrm dB_f}{\mathrm dS}-\tfrac{\mathrm dB_s}{\mathrm dS}\big]\\[-2pt]
&+\mathcal K_{h|f}\big[\tfrac{\mathrm dB_h}{\mathrm dS}-\tfrac{\mathrm dB_f}{\mathrm dS}\big]
\Big],
\end{aligned}
\end{equation}
where $\mathcal K_{c|p}[\eta]$ is the response of the conditional crossing $N(c|p)$ to the bracketed slope profile -- the same functional convolution as \eqref{eq:tid-kernel}, but built on the conditional crossing and its own kernel,
\begin{equation}
\label{eq:Kcp}
\mathcal K_{c|p}[\eta]\equiv\frac{1}{N(c|p)}\int K_{c|p}(S_c,S')\,\eta(S')\,\mathrm dS',
\qquad
K_{c|p}(S_c,S')\equiv\frac{\delta N(c|p)}{\delta B_c(S')}.
\end{equation}
Each bracket in \eqref{eq:tid-nested} is the \emph{excess} barrier slope the walk must clear at that level; the slopes telescope, so the per-level responses sum to the full halo response and the host average reconstructs the direct halo value, the closure \eqref{eq:tid-closure}. This is the tidal analogue of the equivalence between the bias from the conditional mass function \citep{mowhite1996} and the peak--background split \citep{shethtormen1999}: it holds for the full convolution but is violated by the local product \eqref{eq:tid-local}, whose host-averaged and direct values disagree by up to ${\sim}5\times$. Because the conditional response is retained at every level, the tidal sector is a genuine three-level calculation, whereas the density sector of Section~\ref{sec:envbias} collapses to two. Setting the host slopes to zero, $\mathrm dB_f/\mathrm dS=\mathrm dB_s/\mathrm dS=0$, the first two brackets of \eqref{eq:tid-nested} vanish and the per-host response collapses to the conditional term $\tfrac32\,\mathcal K_{h|f}[\mathrm dB_h/\mathrm dS]=\bso^{L}(\Mh|\Mf)$; its average over $P(\Mf|\Mh)$ is the flat-barrier form \eqref{eq:tid-envbias} quoted in the main text.